\documentclass[12pt,draftclsnofoot,onecolumn]{IEEEtran}
\usepackage{amsmath,amssymb,mathtools,dsfont,bm}
\usepackage{algorithmic,algorithm}
\usepackage{cite,graphicx,stfloats,epstopdf,xcolor,url}
\usepackage[mathscr]{eucal}
\usepackage[caption=false,font=footnotesize]{subfig}
\usepackage[colorlinks=false,citecolor=red]{hyperref}
\usepackage{multirow,array,makecell}

\allowdisplaybreaks

\newtheorem{lemma}{Lemma}
\newtheorem{corollary}{Corollary}
\newtheorem{theorem}{Theorem}
\newtheorem{remark}{Remark}

\usepackage{tikz}
\newcommand\copyrightnotice{%
    \begin{tikzpicture}[remember picture,overlay]
    \node[anchor=north,yshift=-7pt] at (current page.north) {\fbox{\parbox{\dimexpr\textwidth-\fboxsep-\fboxrule\relax}{This work has been submitted to the IEEE for possible publication. Copyright may be transferred without notice, after which this version may no longer be accessible.}}};
    \end{tikzpicture}%
}

\begin{document}
\title{The Meta Distribution of SINR in UAV-Assisted Cellular Networks}

\author{Minwei Shi, Kai Yang,~\IEEEmembership{Member,~IEEE}, Dusit Niyato,~\IEEEmembership{Fellow,~IEEE}, \\Hang Yuan, He Zhou, and Zhan Xu
\thanks{M. Shi, K. Yang, and H. Zhou are with the School of Information and Electronics, Beijing Institute of Technology, Beijing 100081, China (email: shiminwei@bit.edu.cn, yangkai@ieee.org, willzhou@bit.edu.cn).

H. Yuan is with the School of Cyberspace Science and Technology, Beijing Institute of Technology, Beijing 100081, China (e-mail: yuanhang@bit.edu.cn).

D. Niyato is with the School of Computer Science and Engineering, Nanyang Technological University, Singapore 639798 (email: dniyato@ntu.edu.sg).

Z. Xu is with the Key Laboratory of Modern Measurement and Control Technology, Beijing Information Science and Technology University, Beijing 100101, China, and also with the School of Information and Communication Engineering, Beijing Information Science and Technology University, Beijing 100101, China (e-mail: xuzhan@bistu.edu.cn).}}

\maketitle
\copyrightnotice

\begin{abstract}
Mounting compact and lightweight base stations on unmanned aerial vehicles (UAVs) is a cost-effective and flexible solution to provide seamless coverage on the existing terrestrial networks. 
While the coverage probability in UAV-assisted cellular networks has been widely investigated, it provides only the first-order statistic of signal-to-interference-plus-noise ratio (SINR). 
In this paper, to analyze high-order statistics of SINR and characterize the disparity among individual links, we provide a meta distribution (MD)-based analytical framework for UAV-assisted cellular networks, in which the probabilistic line-of-sight channel and realistic antenna pattern are taken into account for air-to-ground transmissions. 
To accurately characterize the interference from UAVs, we relax the widely applied uniform off-boresight angle (OBA) assumption and derive the exact distribution of OBA. 
Using stochastic geometry, for both steerable and vertical antenna scenarios, we obtain mathematical expressions for the moments of condition success probability, the SINR MD, and the mean local delay. 
Moreover, we study the asymptotic behavior of the moments as network density approaches infinity. 
Numerical results validate the tightness of the theoretical results and show that the uniform OBA assumption underestimates the network performance, especially in the regime of moderate altitude of UAV. 
We also show that when UAVs are equipped with steerable antennas, the network coverage and user fairness can be optimized simultaneously by carefully adjusting the UAV parameters. 
\end{abstract}


\IEEEpeerreviewmaketitle

\section{Introduction}\label{sec:introduction}

Due to the recent advances in drone manufacturing, mounting compact and lightweight base stations (BSs) on unmanned aerial vehicles (UAVs) has become increasingly feasible \cite{WuXuZengNgAlDh21,XiaoZhuLiuYiZhan22,ZengZhan20BK}. 
Using UAVs as flying BSs can deliver a cost-effective and flexible solution to enhance the coverage and capacity of existing terrestrial cellular networks, due to their capability to adjust altitudes, avoid obstacles, and improve the possibility of line-of-sight (LoS) links to ground users \cite{WuXuZengNgAlDh21}. 
When properly deployed and operated, UAVs are expected to provide reliable and on-demand wireless communications for a variety of real-world scenarios \cite{ZengZhan20BK}.

However, despite such promising opportunities for UAV-BSs, one must address the technical challenges in characterizing UAV-assisted cellular networks, which are due to the distinguishing features in terms of irregular and variable topology, high altitude, as well as the unique air-to-ground (A2G) channel \cite{MozaSaadBennNamDebb19}. 
Taking these features into account, a large number of recent literature has studied the modeling and analysis of UAV-assisted cellular networks using stochastic geometry \cite{ZhouDurrGuoYani19,AlzeYani19,CherAlzeYaniYong21,BanaDhil22,KouzElSaDahrAlshAlNa21,YiLiuBodaNallKara19,_GalkKibiDaSi18,ChenZhan20,_DengChenWei21,MatrKishAlou21,SunWangZhanQuek20,SunDingDai19}. 
Considering a UAV designated to provide coverage for a temporary event, the authors in \cite{ZhouDurrGuoYani19} study the effects of UAV location and A2G channel environment on the downlink and uplink coverage performance in a single UAV-assisted single-cell cellular network. 
By modeling terrestrial BSs (TBSs) and UAVs as two homogeneous Poisson point processes (PPPs), the downlink signal-to-interference-plus-noise ratio (SINR) coverage probabilities for a terrestrial and an aerial user are derived in \cite{AlzeYani19} and \cite{CherAlzeYaniYong21}, respectively. 
More sophisticatedly, the wireless backhauled connection between UAVs and TBSs is considered under the stochastic geometry framework, with relaying protocol in \cite{BanaDhil22}, with backhaul aware transmission scheme in \cite{KouzElSaDahrAlshAlNa21}, and with millimeter-wave backhauling in \cite{YiLiuBodaNallKara19,_GalkKibiDaSi18,KouzElSaDahrAlshAlNa21}. 
Moreover, recent literature has studied the location dependence between TBSs and UAVs \cite{ChenZhan20,_DengChenWei21,MatrKishAlou21}. 
Considering the fact that UAVs are usually circularly deployed around TBSs to avoid mutual interference, the location dependence between TBSs and UAVs is considered in \cite{ChenZhan20,_DengChenWei21}.
In rural area environment, where TBSs are generally getting sparse when moving away from the city center, the authors in \cite{MatrKishAlou21} model the TBSs locations as an inhomogeneous PPP and the UAVs are uniformly distributed outside an exclusion zone.

It is worth noting that the aforementioned literature mainly focuses on the coverage probability, i.e., the complementary cumulative distribution function (CCDF) of SINR. 
While the coverage probability is significant, it characterizes only the \textit{average} performance of the whole network. In other words, it cannot reveal the \textit{variation} of performance among individual links \cite{Haen16,Haen21,DengHaen19a}. 
To address this concern, the authors in \cite{Haen16} establish a framework to evaluate the meta distribution (MD) of signal-to-interference ratio (SIR) in Poisson bipolar and cellular networks, which quantifies the fraction of links that achieve the target SIR threshold above a required reliability level. 
Inspired by its sharpness, the metric of MD is widely analyzed in terrestrial networks \cite{FengHaen20,DengHaen19a,DengHaen17,WangDengWei21,ShiGaoYangNiyaHan21}. 
However, only little literature has investigated this metric in aerial networks except \cite{ZengZhan20BK,PhD2019_PhD2019_Leeds_Hayajneh}, where the MDs of signal-to-noise ratio (SNR) and SIR are evaluated in a single-tier UAV network with Rayleigh fading. 
In the context of UAV-assisted cellular networks, it is even more challenging due to the need for considering the interplay between TBSs and UAVs.

\begin{table}[!t]
    \centering\caption{Comparison Between the Contributions of This Work and Existing Literature Evaluating the Performance of UAV-Assisted Cellular Networks}\label{tab:literature}
    \begin{tabular}{|l|c|c|r|}
        \hline
        \textbf{Reference} & \textbf{Beam Direction} & \makecell[c]{\textbf{OBA Evaluation}} & \textbf{Performance Metric} \\ \hline\hline
        \cite{AlzeYani19,MatrKishAlou21,SunWangZhanQuek20,SunDingDai19} & Isotropic & Accurate & \multirow{5}{*}{\makecell[r]{Coverage probability}} \\ \cline{1-3}
        N.~Cherif~\textit{et al.} \cite{CherAlzeYaniYong21} & \multirow{2}{*}{Vertical} & \multirow{4}{*}{Approximate} &  \\ \cline{1-1}
        M.~Banagar~\textit{et al.} \cite{BanaDhil22} &  &  &  \\ \cline{1-2}
        N.~Kouzayha~\textit{et al.} \cite{KouzElSaDahrAlshAlNa21} & \multirow{2}{*}{Steerable} &  &  \\ \cline{1-1}
        W.~Yi~\textit{et al.} \cite{YiLiuBodaNallKara19} &  &  &  \\ \cline{1-4}
        \textbf{Our contributions} & \makecell[c]{Steerable/Vertical} & \textbf{Accurate} & \textbf{SINR MD} \\ \hline
    \end{tabular}
\end{table}

Another challenge in analyzing UAV-assisted cellular networks lies in the accurate characterization of the aggregate interference from UAVs. 
Specifically, the interference from UAVs depends on the transmission distances and the antenna gains from interfering UAVs to the typical user. The former is determined by the network topology and can be evaluated efficiently. The latter is somewhat complicated to evaluate. 
To elaborate, due to the altitude disparity between terrestrial users and UAVs, users are normally geographically proximate to their target UAVs. 
Thus, from a typical user's perspective, the angle off the boresight of an interfering UAV antenna, which we term the \textit{off-boresight angle (OBA)}, is location-dependent. 
To facilitate the analysis and maintain tractability, most of the literature assumes that the beams of interfering UAVs are randomly oriented with respect to each other so that OBA is \textit{uniformly} distributed in $[0,\pi]$ \cite{KouzElSaDahrAlshAlNa21,YiLiuBodaNallKara19,BanaDhil22,YiLiuDengNall20,LiuHoWu19,AzarGeraGarcPoll20,_ShiDeng21,CherAlzeYaniYong21}. 
This assumption is accurate for terrestrial networks, where the altitudes of BSs and users are comparable. 
However, since the UAVs in the proximity of the typical user are few in quantity yet strong in antenna gain (because of the small OBA), it is non-trivial to judge whether the uniform OBA assumption is accurate in aerial networks and how it affects the network performance in terms of coverage probability and SINR MD.

To address the aforementioned challenges, this paper investigates the SINR MD in a UAV-assisted cellular network, where UAVs and TBSs are modeled as two homogeneous PPPs at different altitudes. 
The unique features of UAVs in terms of the realistic antenna pattern, probabilistic LoS transmission, and Nakagami-$m$ fading are incorporated. 
Using stochastic geometry, we calculate the exact distribution of OBA and derive the SINR MD and mean local delay. 
To the best of the authors' knowledge, this work is the first to investigate SINR MD in UAV-assisted cellular networks and to study OBA distribution. 
The comparison between this work and existing literature evaluating the performance of UAV-assisted cellular networks is sketched in Table~\ref{tab:literature}. 
The contributions of this paper are summarized as follows:
\begin{enumerate}
    \item \textit{Accurate Characterization of A2G Interference:} We relax the widely applied uniform OBA assumption and derive the exact distribution of OBA for the scenarios when UAVs are equipped with steerable or fixed directional antennas. 
    By incorporating the OBA distribution into analysis, we accurately characterize the A2G interference. Numerical results show that the uniform OBA assumption has underestimated the network performance in terms of coverage probability and link reliability, especially in the regime of moderate UAV altitude. 
    \item \textit{Fine-Gained Performance Evaluation:} We provide an MD-based analytical framework for UAV-assisted cellular networks, in which the realistic A2G channels and different UAV antenna patterns are incorporated. 
    We derive the moments of conditional success probability (CSP), based on which the mean local delay and the tightly approximated SINR MD are obtained. 
    {To obtain more insights into the ultra-dense networks, we further explore the asymptotic behavior of the association probability and the CSP moments. Several relevant special cases are also introduced for possible simplification of analytical results.} 
    \item \textit{Design Guidelines and Insights:} We validate the above analytical results using extensive simulations and show the necessity of considering the OBA distribution when UAVs are equipped with directional antennas. 
    We reveal the superiority of equipping steerable antennas at UAVs in network coverage and user fairness enhancement and show the potential advantage of replacing TBSs with UAVs in the proposed framework. 
\end{enumerate}

The rest of the paper is organized as follows. In Section~\ref{sec:system_model}, we introduce the system model. Then, we analyze the association probability and the exact distribution of OBA in Section~\ref{sec:association_analysis}. Based on these results, in Section~\ref{sec:meta_distribution}, we derive the metrics in terms of the $b$-th moment of CSP, SINR MD, and mean local delay, followed by the discussions of asymptotic behavior and several special cases. We provide simulation results and investigate the impacts of UAV features on the fine-grained network performance in Section~\ref{sec:simulation}. Finally, we conclude in Section~\ref{sec:conclusion}. For the convenience of presentation, we list the main notations in Table~\ref{tab:parameter}.

\begin{table}[!t]
    \centering\caption{Notations and Descriptions}\label{tab:parameter}
    \begin{tabular}{|p{60pt}|p{350pt}|}
        \hline
        \bfseries Notation & \bfseries Description\\
        \hline\hline
        $\Phi_k$, $\lambda_k$ & Set of the $k$-th tier transmitters and its density, $k\in\{\mathrm{b},\mathrm{L},\mathrm{N}\}$\\
        \hline
        $\bar{\Phi}_k$ & Distance point process of the $k$-th tier transmitters\\
        \hline
        $\bar{\lambda}_k$, $\bar{\Lambda}_k$ & Density and intensity measure of $\bar{\Phi}_k$\\
        \hline
        $R_{0,k}$ & Minimum distance from the typical user to the $k$-th tier transmitters\\
        \hline
        $Y_{0,k}$ & Serving link distance when the typical user is associated with the $k$-th tier\\
        \hline
        $L$ & Horizontal distance between the typical user and the projection of its serving UAV\\
        \hline
        $P_k$ & Transmit power of the $k$-th tier\\
        \hline
        $p_{\mathrm{L}}(r)$, $p_{\mathrm{N}}(r)$ & LoS and NLoS probabilities of an A2G link with distance $r$\\
        \hline
        $\delta(X)$ & Indicator of the LoS/NLoS condition for the link from $X$ to the typical user\\
        \hline
        $\alpha_k$, $\kappa_k$ & Path loss exponent and intercept of the link from the $k$-th tier to the typical user\\
        \hline
        $\mathsf{M}_k$ & Small-scale fading parameter of the $k$-th tier\\
        \hline
        $\mathsf{H}_{X}$, $\mathsf{H}_{k,j}$ & Small-scale fading gain from $X$ and $X_{k,j}$ to the typical UE\\
        \hline
        $G_{\mathrm{M},k}$ & Maximum antenna gain of the $k$-th tier transmitters\\
        \hline
        $G_k(\theta)$ & Antenna gain of the $k$-th tier transmitter with angle $\theta$ off its boresight direction\\
        \hline
        $\vartheta_{3\mathrm{dB}}$, $\mu _{\mathrm{SLA}}$ & The $3$\,dB beamwidth and the sidelobe attenuation limit of UAVs
        \\
        \hline
        $\mathcal{P}_\mathrm{s}(\gamma)$ & CSP with SINR threshold $\gamma$\\
        \hline
        $M_b(\gamma)$ & The $b$-th moment of $\mathcal{P}_\mathrm{s}(\gamma)$\\
        \hline
        $A_k$ & Association probability of the $k$-th tier\\
        \hline
        $\gamma$ & SINR threshold\\
        \hline
        ${N_0}$ & Thermal noise power\\
        \hline
        $\mathbb{P}_X[\cdot]$; $\mathbb{E}_X[\cdot]$ & Probability of an event; Expectation with respect to the random variable $X$\\
        \hline
        $f_X(x)$, $F_X(x)$, $\bar{F}_X(x)$ & The probability density function (PDF), cumulative distribution function (CDF), and CCDF of $X$, respectively\\
        \hline
        $\lVert\cdot\rVert$; $\overline{AB}$ & Euclidean metric on $\mathbb{R}^2$; Length of the line segment $AB$\\
        \hline
        ${}_2F_1$; $\mathbf{1}_{A}\left( \cdot \right)$ & Gaussian hypergeometric function; Indicator function of $A$\\
        \hline
    \end{tabular}
\end{table}

\section{System Model}\label{sec:system_model}

\subsection{Spatial Setup}\label{subsec:2A-spatial}

\begin{figure}[!t]
    \centering
    \includegraphics[width=0.6\textwidth]{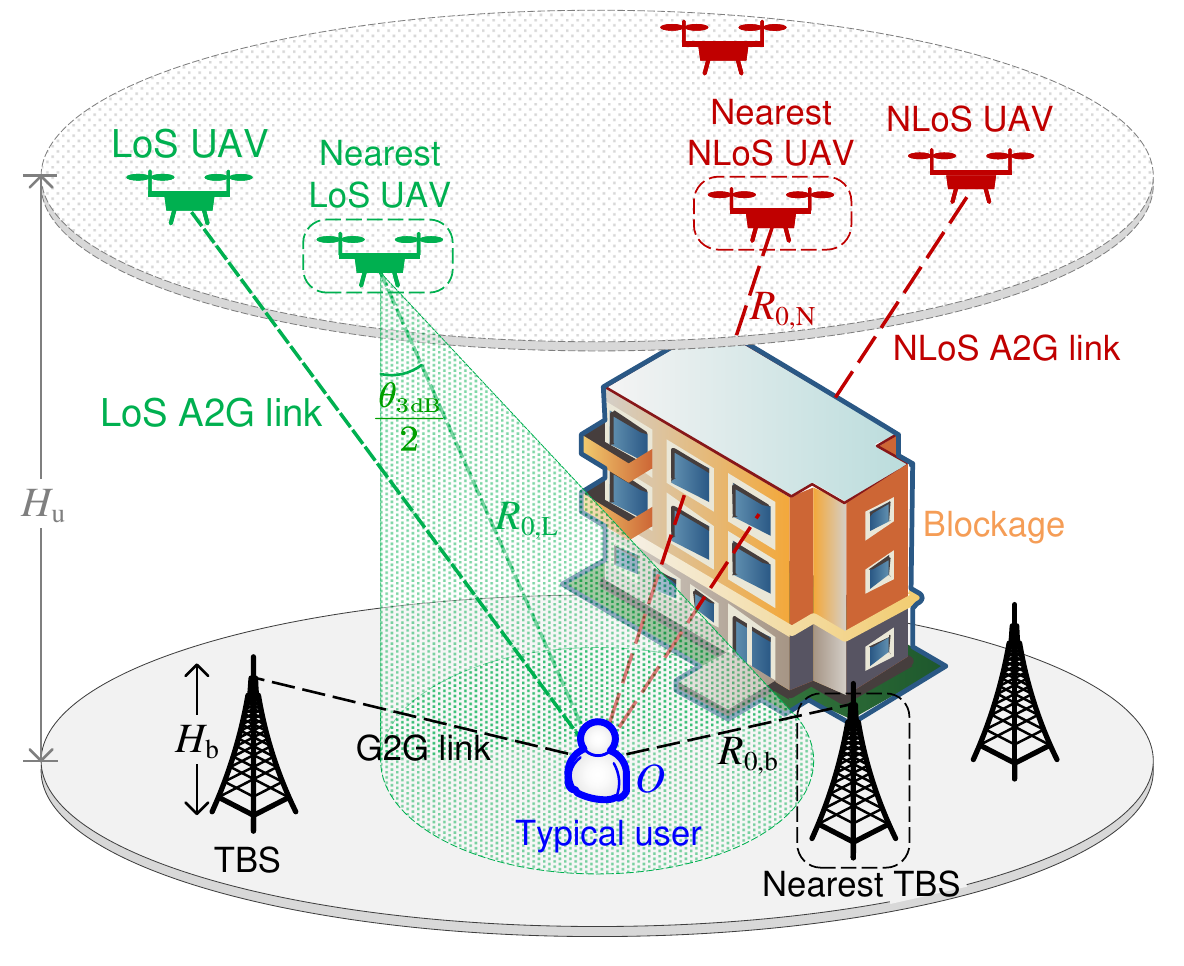}
    \caption{Illustration of the network geometry and parameters of UAV altitude $H$, antenna beamwidth $\theta_{\mathrm{3dB}}$, and the minimum distance $R_{0,k}$, $k\in\mathcal{K}$. Here UAVs are equipped with steerable antennas and the typical user is currently associated with the nearest LoS UAV.}
    \label{fig:1}
\end{figure}

We consider the downlink of UAV-assisted cellular networks as illustrated in Fig.~\ref{fig:1}. 
{The TBSs are deployed at an altitude $H_{\mathrm{b}}$ following a homogeneous PPP $\Phi _{\mathrm{b}}=\left\{ X_{\mathrm{b},1},X_{\mathrm{b},2},\ldots \right\}$ with density $\lambda _{\mathrm{b}}$. 
Independently from $\Phi _{\mathrm{b}}$, the UAVs $\Phi _{\mathrm{u}}=\left\{ X_{\mathrm{u},1},X_{\mathrm{u},2},\ldots \right\}$ hover at a fixed altitude $H_{\mathrm{u}}$ and follow a homogeneous PPP with density $\lambda _{\mathrm{u}}$\footnote{The analysis of spatial models with inter-tier dependence is beyond the scope of this paper and is left to the future work.}.} 
The TBSs and UAVs are also assumed to operate in the same frequency spectrum and transmit with power $P_{\mathrm{b}}$ and $P_{\mathrm{u}}$, respectively. 
Perfect wireless backhaul between TBSs and UAVs is considered so that they can exchange information accurately \cite{AlzeYani19,MatrKishAlou21,SunDingDai19}.
{Besides, users are independently deployed on the ground\footnote{Although the strategic placement of UAVs is possible to optimize the network utility \cite{SenaDurrZhouYangDing20}, in the absence of exact traffic and blockage patterns, these optimal locations are not known, which justifies the assumption that UAVs are randomly deployed and are not necessarily located over their serving users \cite{ChetDhil17}.}.} 
Each TBS/UAV is assumed to have a terrestrial user to serve so that the considered network is fully loaded. 
Without loss of generality, we conduct our analysis on the typical user located at the origin $O$ \cite{BaccBlas09BK}.

\subsection{Propagation Model}\label{subsec:2B-propagation}

\subsubsection{LoS Model}

One of the main advantages of UAVs is establishing LoS links to terrestrial users, which can significantly improve the signal quality compared to non-LoS (NLoS) links. 
In this paper, we consider the probabilistic LoS model for A2G links. Specifically, the LoS probability of an A2G link with distance $r$ is given by \cite{AlHoKandLard14}
\begin{align}
    p_{\mathrm{L}}\left( r \right) =\frac{1}{1+\mu _a\exp \left\{ -\mu _b\left[ \frac{180}{\pi}\arcsin \left( \frac{H_{\mathrm{u}}}{r} \right) -\mu _a \right] \right\}},\label{eq:los-fun-sigmoid}
\end{align}
where $\mu_{a}$ and $\mu_{b}$ are positive parameters related to the environment and can be obtained by fitting \eqref{eq:los-fun-sigmoid} with the International Telecommunications Union (ITU)-recommended LoS probability function. The NLoS probability of an A2G link with distance $r$ is $p_{\mathrm{N}}\left( r \right)=1-p_{\mathrm{L}}\left( r \right)$. 
From the perspective of the typical user, $\Phi_{\mathrm{u}}$ can be divided into LoS UAVs $\Phi _{\mathrm{L}}$ and NLoS UAVs $\Phi _{\mathrm{N}}$. Here $\Phi _{\mathrm{L}}$ and $\Phi _{\mathrm{N}}$ are PPPs with intensities $\lambda _{\mathrm{u}}p_{\mathrm{L}}\left( r \right)$ and $\lambda _{\mathrm{u}}p_{\mathrm{N}}\left( r \right)$, respectively. 
Besides, since the altitude of TBS is usually comparable to that of buildings in urban environments, we assume that the ground-to-ground (G2G) links between TBSs and users are always in NLoS.

\subsubsection{Path-Loss and Fading}

Denote the set of possible link conditions by $\mathcal{K} \coloneqq \left\{ \mathrm{b},\mathrm{L},\mathrm{N} \right\}$, where $\mathrm{b}$, $\mathrm{L}$, and $\mathrm{N}$ stand for G2G, A2G LoS, and A2G NLoS, respectively. 
For the communication links of condition $k\in\mathcal{K}$, the channel power gain between transmitter $X$ and typical user $O$ consists of large-scale path-loss $L_k\left( r \right)$ and small-scale fading $\mathsf{H}_k$. 
In this paper, we consider the power-law path-loss function $L_k\left( r \right) =\kappa _kr^{-\alpha _k},k\in \mathcal{K}$, where $\alpha _k$ and $\kappa _k$ are path-loss exponent and intercept, respectively. 
Furthermore, $\mathsf{H}_k$ is assumed to follow normalized Gamma distribution with parameter $\mathsf{M} _k$, i.e., $\mathsf{H}_k\sim \mathrm{Gamma}\left( \mathsf{M} _k,1/\mathsf{M} _k \right), k\in \mathcal{K}$. Particularly, we consider Rayleigh fading for G2G links and thus $\mathsf{M}_{\mathrm{b}}=1$.

\subsection{Antenna Pattern}\label{subsec:2C-antenna}

{We consider that TBSs and UAVs are equipped with downtilted directional antennas to establish access links with ground users. 
Based on the third generation partnership project (3GPP) specification \cite{3GPP_TR_36p873}, the antenna gain along the direction of angle $\theta$ off the boresight of a TBS ($j=\mathrm{b}$) or a UAV ($j=\mathrm{u}$) can be represented by 
\begin{align}
    G_j\left( \theta \right) =G_{\mathrm{M},j}\cdot 10^{-\min \left\{ 12\left( \theta /\vartheta _{3\mathrm{dB},j} \right) ^2,\,\mu _{\mathrm{SLA}} \right\}},\quad \theta \in \left[ 0,\pi \right] ,
\end{align}
where $G_{\mathrm{M},j}$ is the maximum gain, $\vartheta _{3\mathrm{dB},j}$ is the $3$\,dB beamwidth, and $\mu _{\mathrm{SLA}}=20$\,dB is the sidelobe attenuation limit. 
As a special case, when the target is exactly in the direction of the antenna's boresight, we have $\theta=0$ and $G_{j}\left( 0 \right) =G_{\mathrm{M},j}$.}

{Regarding the antenna directions, for TBS antennas, their mainlobes are considered to be tilted downward to the ground. On the other hand, for UAV antennas, we consider both \textit{steerable} and \textit{vertical} scenarios.} To be specific, in the steerable antenna scenario, one UAV is able to adjust its antenna steering orientation, mechanically or electrically, toward its intended user exactly, to maximize the directivity gain \cite{BanaDhil22}. 
In the vertical antenna scenario, the mainlobe direction is static and tilts vertically toward the ground \cite{BanaDhil22,_DengChenWei21,SaleHoss19a}. 
Hereafter we use $\mathrm{SA}$ and $\mathrm{VA}$ to indicate ``steerable antenna'' and ``vertical antenna'', respectively.

{Consequently, in $\varsigma\in\{\mathrm{SA},\mathrm{VA}\}$ scenario, the \textit{average} power received from a \textit{serving} TBS/UAV in the $k$-th tier is expressed as 
\begin{align}
    l_{k}^{\left( \varsigma \right)}\left( r \right) =\begin{cases}
        P_{\mathrm{b}}G_{\mathrm{b}}\left( \arccos \left( H_{\mathrm{b}}/r \right) \right) \kappa _{\mathrm{b}}r^{-\alpha _{\mathrm{b}}},&     k=\mathrm{b},\\
        P_{\mathrm{u}}G_{\mathrm{u}}\left( \arccos \left( H_{\mathrm{u}}/r \right) \right) \kappa _kr^{-\alpha _k},&       k\in \left\{ \mathrm{L},\mathrm{N} \right\} ,\varsigma =\mathrm{VA},\\
        P_{\mathrm{u}}G_{\mathrm{u}}\left( 0 \right) \kappa _kr^{-\alpha _k},&      k\in \left\{ \mathrm{L},\mathrm{N} \right\} ,\varsigma =\mathrm{SA},\\
    \end{cases}\label{eq:l_k}
\end{align}
where $r$ is the link distance.}

\begin{remark}
Although we adopt 3GPP-based antenna pattern, the subsequent analyses in Sections~\ref{sec:association_analysis} and \ref{sec:meta_distribution} can be straightforwardly generalized to arbitrary symmetric antenna pattern.
\end{remark}

\subsection{Association Strategy}\label{subsec:2D-strategy}

{Due to the random fading effects, any transmitter can potentially have the best channel condition with the typical user \cite{BehnWang15,BehnBeau15}. 
However, considering the path-loss induced signal attenuation, this serving transmitter is more likely to be the one that provides the \textit{strongest average received power} \cite{LiuWangChenElkaWong16}. 
In this paper, the typical user is associated with the transmitter that provides the strongest average received power \cite{BanaDhil22}. 
Since the small-scale fading gain $\mathsf{H}_k$ is averaged out and $\mathbb{E} \left[ \mathsf{H}_k \right] =1$, the serving transmitter can be formulated by 
\begin{align}
    {X^{\star}=\underset{X\in \cup _{k\in \mathcal{K}}\Phi _k}{\mathrm{arg}\max}l_{\delta \left( X \right)}^{\left( \varsigma \right)}\left( \left\| X \right\| \right),}
\end{align}
where $\delta \left( X \right) \coloneqq \left\{ k\in \mathcal{K} :X\in \Phi _k \right\}$ indicates the link condition of $X\rightarrow O$. 
Note that $X^{\star}$ can be a TBS, LoS UAV, or NLoS UAV as long as it is near enough to the typical user.} 
{When the typical user is associated with the $k$-th tier, i.e., $\delta \left( X^{\star} \right) =k$, the downlink received SINR can be written as 
\begin{align}
    \mathrm{SINR}=\frac{l_{k}^{\left( \varsigma \right)}\left( \left\| X^{\star} \right\| \right) \mathsf{H}_{X^{\star}}}{N_0+\sum_{\ell \in \mathcal{K}}{I_{\ell}}}, \label{eq:sinr_define}
\end{align}
where $I_{\ell}$ denotes the interference constituted from tier $\ell$, i.e., 
\begin{align}
    I_{\ell}=\sum_{X\in \Phi _{\ell}\backslash \left\{ X^{\star} \right\}}{P_{\ell}G_X\mathsf{H}_X\kappa _{\ell}\left\| X \right\| ^{-\alpha _{\ell}}}. \label{eq:Ik_define}
\end{align}}

\begin{remark}\label{remark:2}
Due to the different path-loss parameters and transmit powers in each tier, the serving TBS/UAV is not necessarily the nearest one to the user. 
{For instance, the average received power from a farther UAV to the typical user may be stronger than that from a nearer UAV.} 
However, when restricted to a specific set $\Phi_k$, $k\in\mathcal{K}$, where the path-loss parameters and transmit powers are the same, the closest one will provide the strongest average received power in $\Phi_k$. 
Therefore, the candidates of serving transmitters are the closest TBS, LoS UAV, or NLoS UAV. 
{The probability of the typical user being associated with each tier will be derived in the next section.}
\end{remark}

\section{Association and OBA Analysis}\label{sec:association_analysis}

In this section, we provide some important intermediate results that help to analyze the SINR MD in UAV-assisted cellular networks. Firstly, we derive the point process intensity for the distances from the transmitters in each tier in Section~\ref{subsec:3A-distance}, based on which the association probability of each tier is obtained in Section~\ref{subsec:3B-association}. Then, for both $\mathrm{SA}$ and $\mathrm{VA}$ scenarios, we derive the exact distribution of OBA in Section~\ref{subsec:3C-angle}.

\subsection{Distance Point Process}\label{subsec:3A-distance}

According to Section~\ref{subsec:2B-propagation}, observed from the typical user, the set of transmitters is divided into three subsets according to the G2G/A2G and LoS/NLoS statuses of the communication links, i.e., $\Phi_{\mathrm{b}}\cup\Phi_{\mathrm{u}}=\cup_{k\in\mathcal{K}}\Phi_k$. 
The transmitters in $\Phi_k$ constitute a three-dimensional non-homogeneous PPP. 
For ease of analysis, we map $\Phi_k$ to a one-dimensional point process $\bar{\Phi}_k\coloneqq \left\{ \left\| X \right\| :X\in \Phi _k \right\}$, which represents the distances from the points in $\Phi_k$ to the origin. 
From the mapping theorem of PPP \cite{BlasHaenKeelMukh18BK}, we know that $\bar{\Phi}_k$ is still Poisson. 
The following lemma presents the intensity measure $\bar{\Lambda}_k\left( \left[ 0,r \right] \right)$ and the intensity $\bar{\lambda}_k\left( r \right)$ of distance point process $\bar{\Phi}_k$.

\begin{lemma}[Distance Point Process for UAV-Assisted Cellular Network]\label{lemma:distance_pp_intensity}
For $k\in\mathcal{K}$, the intensity measure of $\bar{\Phi}_k$ is given as
\begin{align}\label{eq:lemma1_Lambda}
    \bar{\Lambda}_k\left( \left[ 0,r \right] \right) =\begin{cases}
        \pi \lambda _{\mathrm{b}}\left( r^2-H_{\mathrm{b}}^2 \right), & r\ge H_{\mathrm{b}},k=\mathrm{b},\\
        2\pi \lambda _{\mathrm{u}}\int_{H_{\mathrm{u}}}^r{p_k\left( x \right) x\,\mathrm{d}x}, & r\ge H_{\mathrm{u}},k\in \left\{ \mathrm{L},\mathrm{N} \right\},\\
        0, & \mathrm{otherwise},\\
    \end{cases}
\end{align}
and the intensity of $\bar{\Phi}_k$ is
\begin{align}\label{eq:lemma1_lambda}
    \bar{\lambda}_k\left( r \right) =\begin{cases}
        2\pi \lambda _{\mathrm{b}}r, & r\ge H_{\mathrm{b}},k=\mathrm{b},\\
        2\pi \lambda _{\mathrm{u}}rp_k\left( r \right), & r\ge H_{\mathrm{u}},k\in \left\{ \mathrm{L},\mathrm{N} \right\},\\
        0, & \mathrm{otherwise}.\\
    \end{cases}
\end{align}
\end{lemma}
\begin{IEEEproof}
According to the definition of intensity measure, we have $\bar{\Lambda}_k\left( \left[ 0,r \right] \right) =\mathbb{E} \left[ \Phi _k\left( \mathcal{B} \left( O,r \right) \right) \right]$ \cite{BlasHaenKeelMukh18BK}, where $\mathcal{B} \left( O,r \right)$ denotes a ball with center $O$ and radius $r$ and $\Phi _k\left( \cdot \right)$ is viewed as a counting measure. 
Since $\Phi_k$ is a PPP with intensity known in Section~\ref{subsec:2B-propagation}, $\bar{\Lambda}_k\left( \left[ 0,r \right] \right)$ can be easily derived by evaluating the expectation of $\Phi _k\left( \mathcal{B} \left( O,r \right) \right)$. The intensity $\bar{\lambda}_k\left( r \right)$ is then obtained by taking the derivative of $\bar{\Lambda}_k\left( \left[ 0,r \right] \right)$ with respect to $r$.
\end{IEEEproof}

As we have stated in Remark~\ref{remark:2}, the associated tier is dominated by the minimum distance of each tier. 
We let $R_{0,k}$ denote the minimum distance from the typical user to $\Phi_k$, i.e., $R_{0,k}\coloneqq \min \,\bar{\Phi}_k=\min \left\{ \left\| X \right\| :X\in \Phi _k \right\}$. The distribution of $R_{0,k}$ is presented in the following lemma.

\begin{lemma}[Minimum Distance Distribution]\label{lemma:r0k_cdf_pdf}
For the typical user in UAV-assisted cellular network, the CDF and PDF of the distance to the nearest transmitter in $\Phi_k$ are given by
\begin{align}
    F_{R_{0,k}}\left( r \right) &=1-e^{-\bar{\Lambda}_k\left( \left[ 0,r \right] \right)},\quad r\ge 0,\label{eq:r0k_cdf}
    \\
    f_{R_{0,k}}\left( r \right) &=\bar{\lambda}_k\left( r \right) e^{-\bar{\Lambda}_k\left( \left[ 0,r \right] \right)},\quad r\ge 0,\label{eq:r0k_pdf}
\end{align}
respectively.
\end{lemma}
\begin{IEEEproof}
The proof can be directly obtained from the void probability of PPPs \cite{BaccBlas09BK}.
\end{IEEEproof}

\subsection{Association Probability}\label{subsec:3B-association}

Based on the results in Section~\ref{subsec:3A-distance}, we obtain the association probability for each combination of antenna type and channel condition in the following theorem.

\begin{theorem}[Association Probability]\label{theo:asso_prob}
When UAVs are equipped with antennas of type $\varsigma\in\{\mathrm{SA},\mathrm{VA}\}$, the probability that the typical user is associated with the $k$-th tier is given as
\begin{align}
    A_k^{(\varsigma)}=\int_0^{\infty}{\bar{\lambda}_k\left( r \right) e^{-\sum_{\ell \in \mathcal{K}}{\bar{\Lambda}_{\ell}\left( \left[ 0,\chi _{k,\ell}^{(\varsigma)}\left( r \right) \right] \right)}}\,\mathrm{d}r},~~k\in \mathcal{K},~~\varsigma\in\{\mathrm{SA},\mathrm{VA}\},\label{eq:asso_prob}
\end{align}
{where $\chi _{k,\ell}^{\left( \varsigma \right)}\left( r \right) \coloneqq \dot{l}_{\ell}^{(\varsigma)}(l_{k}^{(\varsigma)}(r))$ and $\dot{l}_{\ell}^{\left( \varsigma \right)}\left( \cdot \right)$ denotes the inverse function of $l_{k}^{\left( \varsigma \right)}(\cdot)$.} 
\end{theorem}
\begin{IEEEproof}
See Appendix~\ref{app:theo_asso_prob}.
\end{IEEEproof}

With the condition that the typical user is associated with the $k$-th tier, we let $Y_{0,k}$ denote the distance of the \textit{serving link}. The following corollary presents the distribution of $Y_{0,k}$.

\begin{corollary}[Serving Link Distance Distribution]\label{coro:y0k_pdf}
When UAVs are equipped with antennas of type $\varsigma\in\{\mathrm{SA},\mathrm{VA}\}$ and the typical user is associated with the $k$-th tier, the PDF of the serving link distance is
\begin{align}
    f_{Y_{0,k}^{\left( \varsigma \right)}}\left( r \right) =\frac{\bar{\lambda}_k\left( r \right)}{A_{k}^{\left( \varsigma \right)}}e^{-\sum_{\ell \in \mathcal{K}}{\bar{\Lambda}_{\ell}\left( \left[ 0,\chi _{k,\ell}^{\left( \varsigma \right)}\left( r \right) \right] \right)}},\quad r\ge 0,\label{eq:y0k_pdf}
\end{align}
where $\chi _{k,\ell}^{(\varsigma)}\left( r \right)$ is defined in \textbf{Theorem~\ref{theo:asso_prob}}.
\end{corollary}
\begin{IEEEproof}
According to the conditional probability formula, we have
\begin{align}
    F_{Y_{0,k}^{\left( \varsigma \right)}}\left( y \right) &=\mathbb{P} \left( R_{0,k}\le y \middle|X^{\star}\in\Phi_k \right)\notag
    \\
    &=\frac{\mathbb{P} \left( R_{0,k}\le y,X^{\star}\in \Phi _k \right)}{\mathbb{P} \left( X^{\star}\in \Phi _k \right)}\notag
    \\
    &=\frac{1}{A_{k}^{\left( \varsigma \right)}}\int_0^y{f_{R_{0,k}}\left( r \right) \prod_{\ell \in \mathcal{K} \backslash \left\{ k \right\}}{\bar{F}_{R_{0,\ell}}\left( \dot{l}_{\ell}^{\left( \varsigma \right)}\left( l_{k}^{\left( \varsigma \right)}\left( r \right) \right) \right)}\,\mathrm{d}r}.\label{eq:proof_y0k_cdf}
\end{align}
Then \eqref{eq:y0k_pdf} is obtained by taking the derivative of \eqref{eq:proof_y0k_cdf} with respect to $y$.
\end{IEEEproof}

\subsection{OBA Distribution}\label{subsec:3C-angle}

\begin{figure}[!t]
    \centering
    \subfloat[$\mathrm{SA}$ scenario]{\includegraphics[width=0.5\textwidth]{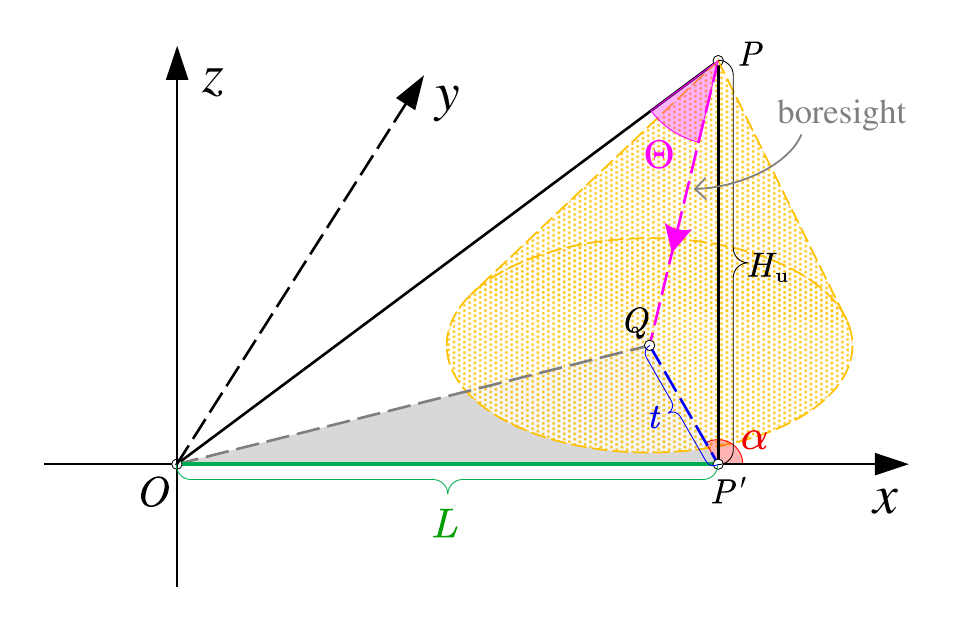}\label{fig:2a}}
    \hfil
    \subfloat[$\mathrm{VA}$ scenario]{\includegraphics[width=0.5\textwidth]{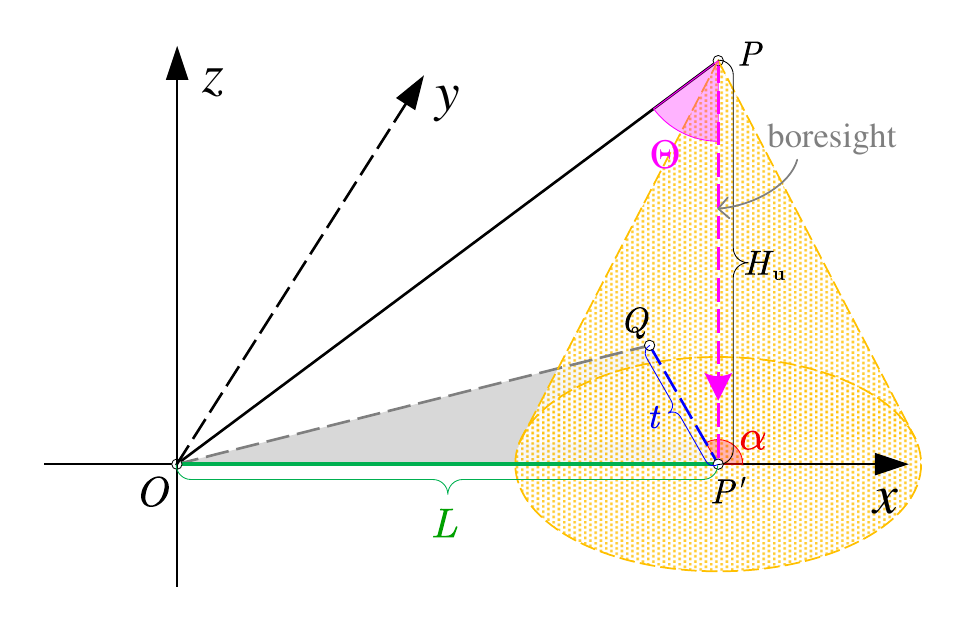}\label{fig:2b}}
    \caption{Illustrations of OBA $\Theta$ for $\mathrm{SA}$ and $\mathrm{VA}$ scenarios. Here $O$ is the typical user. The interfering UAV $P$, with projection $P^\prime$ on the $xy$ plane, is serving a ground user $Q$. The antenna boresights for $\mathrm{SA}$ and $\mathrm{VA}$ scenarios are $PQ$ and $PP^{\prime}$, respectively.}
    \label{fig:2}
\end{figure}

We now focus on the distribution of OBA $\Theta$. 
In Fig.~\ref{fig:2}, we consider a situation where the terrestrial users $Q$ and $O$ are associated with UAV $P$ other TBS/UAV, respectively. 
Thus from the perspective of the typical user $O$, $P$ acts as an interfering UAV.
In order to calculate the interference power received from $P$, we need to determine the antenna gain from $P$, which is related to the angle between $\overrightarrow{PO}$ and the boresight of UAV $P$.

Without loss of generality, we denote the coordinate of $P$ by $(L,0,H_{\mathrm{u}})$. 
Thus $P^{\prime}$, the projection of $P$ on the ground, has coordinate $(L,0,0)$. 
Besides, we let $t$ denote the horizontal distance between $P$ and $Q$ and let $\alpha$ denote the angle between $\overrightarrow{P^{\prime}Q}$ and the positive $x$-axis, i.e., $t =\overline{P^\prime Q}$ and $\alpha =\pi -\angle OP^\prime Q$. 
According to the description in Section~\ref{subsec:2C-antenna}, when equipped with antennas of type $\mathrm{SA}$ and $\mathrm{VA}$, UAV $P$ will point its beam toward $Q$ and $P^{\prime}$, respectively. 
Consequently, the OBA $\Theta$ is equivalent to $\angle OPQ$ and $\angle OPP^\prime$ for $\mathrm{SA}$ and $\mathrm{VA}$ scenarios, respectively. 
The following lemma presents the distribution of $\Theta$.

\begin{lemma}\label{lemma:theta_cdf_pdf_cond_l_r}
{Let $t$ and $L$ denote the horizontal distances from the interfering UAV to its target user and the typical user, respectively.} {For the $\mathrm{SA}$ scenario, the CDF and PDF of $\Theta$ are given as 
\begin{align}
    F_{\left. \Theta \right|L,t}\left( \theta \right) &=
    \begin{cases}
        0, & \theta \in \left[ 0,\theta _{\min} \right) ,\\
        1-\frac{1}{\pi}\arccos \left( \frac{H_{\mathrm{u}}^2- \sqrt{\left( H_{\mathrm{u}}^2+L^2 \right) \left( H_{\mathrm{u}}^2+t^2 \right)}\cos \theta}{Lt} \right), & \theta \in \left[ \theta _{\min} ,\theta _{\max} \right) ,\\
        1, & \theta \in \left[ \theta _{\max} ,\pi \right] ,\\
    \end{cases}\label{eq:theta_cdf_cond_l_r}
    \\
    f_{\left. \Theta \right|L,t}\left( \theta \right) &=
    \begin{cases}
        \displaystyle\frac{\sin \theta}{\pi \sqrt{\left( \cos \theta _{\min} -\cos \theta \right) \left( \cos \theta -\cos \theta _{\max} \right)}}, & \theta \in \left( \theta _{\min} ,\theta _{\max} \right) ,\\
        0, & \mathrm{otherwise},\\
    \end{cases}\label{eq:theta_pdf_cond_l_r}
\end{align}
respectively, where $\theta _{\min} =\arccos \Big( \frac{H_{\mathrm{u}}^2+Lt}{\sqrt{\left( H_{\mathrm{u}}^2+L^2 \right) \left( H_{\mathrm{u}}^2+t^2 \right)}} \Big)$ and $\theta _{\max} =\arccos \Big( \frac{H_{\mathrm{u}}^2-Lt}{\sqrt{\left( H_{\mathrm{u}}^2+L^2 \right) \left( H_{\mathrm{u}}^2+t^2 \right)}} \Big)$.} 
For the $\mathrm{VA}$ scenario, $\Theta \equiv \arctan \left( L/H_{\mathrm{u}} \right)$. 
\end{lemma}
\begin{IEEEproof}
See Appendix~\ref{app:lemma_theta_cdf_pdf_cond_l_r}.
\end{IEEEproof}

Note that $\overline{P^\prime Q}$ is determined by the UAV altitude and the random serving link distance of tier $k\in\mathcal{K}\backslash\{\mathrm{b}\}$. By deconditioning $f_{\left. \Theta \right|L,t}\left( \theta \right)$ over $t$, we obtain the following lemma.

\begin{lemma}[Conditional Distribution of $\Theta$]\label{lemma:theta_cdf_pdf_cond_l}
Conditioned on the horizontal distance between the typical user and interfering UAV being $L$, for the $\mathrm{SA}$ scenario, the PDF of $\Theta$ is given as 
\begin{align}
    f_{\left. \Theta \right|L}\left( \theta \right) =\int_{H_{\mathrm{u}}}^{\infty}{\frac{f_{\left. \Theta \right|L,\sqrt{y^2-H_{\mathrm{u}}^2}}\left( \theta \right) \sum_{k\in \mathcal{K} \backslash \left\{ \mathrm{b} \right\}}{A_{k}^{\left( \varsigma \right)}f_{Y_{0,k}}\left( y \right)}}{1-A_{\mathrm{b}}^{\left( \varsigma \right)}}\,\mathrm{d}y},\quad\theta \in \left[ 0,\pi \right] .
\end{align}
For the $\mathrm{VA}$ scenario, $\Theta \equiv \arctan \left( L/H_{\mathrm{u}} \right)$. 
\end{lemma}
\begin{IEEEproof}
The proof is completed by taking the expectation of $f_{\left. \Theta \right|L,t}\left( \theta \right)$ with respect to $t$, i.e., $f_{\left. \Theta \right|L}\left( \theta \right) =\mathbb{E} _{Y_0}\left[ f_{\left. \Theta \right|L,\sqrt{Y_{0}^{2}-H_{\mathrm{u}}^2}}\left( \theta \right) \right]$. 
\end{IEEEproof}

Based on the conditional distribution of $\Theta$, we directly obtain the average antenna gain from an interfering UAV with horizontal distance $L$ in the following corollary.

\begin{corollary}[Mean of Interfering Antenna Gain]\label{coro:average-gain}
Conditioned on the distance between the typical user and the interfering UAV being $r$, the mean of interfering antenna gain is given as
\begin{align}
    \bar{g}_{\mathrm{I},\mathrm{u}}^{\left( \varsigma \right)}\left( r \right) =\begin{cases}
        \int_0^{\pi}{G_{\mathrm{u}}\left( \theta \right) f_{\Theta |\sqrt{r^2-H_{\mathrm{u}}^2}}\left( \theta \right) \,\mathrm{d}\theta}, & \varsigma =\mathrm{SA},\\
        G_{\mathrm{u}}\left( \arccos \left( H_{\mathrm{u}}/r \right) \right), & \varsigma =\mathrm{VA}.\\
    \end{cases}
\end{align}
\end{corollary}
\begin{IEEEproof}
The proof is straightforward by taking the expectation of $G_{\mathrm{u}}(\theta)$ with respect to $\theta$ and is thus omitted here.
\end{IEEEproof}

{Consequently, in $\varsigma\in\{\mathrm{SA},\mathrm{VA}\}$ scenario, the average power received from an \textit{interfering} TBS/UAV of distance $r$ in the $k$-th tier is expressed as 
\begin{align}
    h_{k}^{\left( \varsigma \right)}\left( r \right) =\begin{cases}
        P_{\mathrm{b}}G_{\mathrm{b}}\left( \arccos \left( H_{\mathrm{b}}/r \right) \right) \kappa _{\mathrm{b}}r^{-\alpha _{\mathrm{b}}},&     k=\mathrm{b},\\
        P_{\mathrm{u}}\bar{g}_{\mathrm{I},\mathrm{u}}^{\left( \varsigma \right)}\left( r \right) \kappa _kr^{-\alpha _k},&      k\in \left\{ \mathrm{L},\mathrm{N} \right\} .\\
    \end{cases}\label{eq:h_k}
\end{align}}
{Recalling \eqref{eq:Ik_define}, the average interference from the $k$-th tier is $\mathbb{E} \left[ I_k \right] =\sum_{X\in \Phi _k\backslash \left\{ X^{\star} \right\}}{h_{k}^{\left( \varsigma \right)}\left( \left\| X \right\| \right)}$.} 

\begin{figure}[!t]
    \centering
    \subfloat[PDF of OBA when $L=50$\,m]{\includegraphics[width=0.5\textwidth]{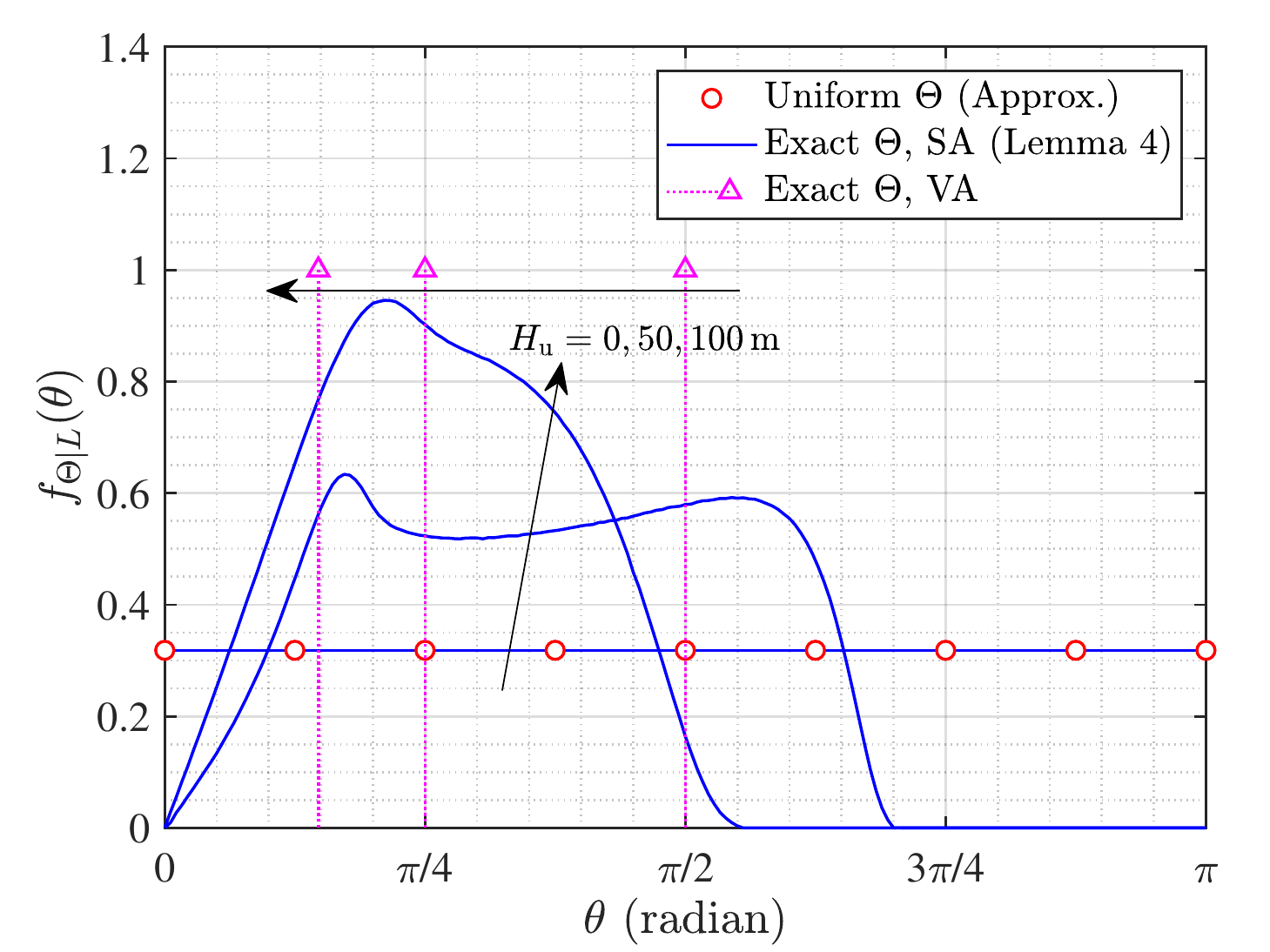}\label{fig:3a}}\hfil
    \subfloat[Average antenna gain when $H_{\mathrm{u}}=100$\,m]{\includegraphics[width=0.5\textwidth]{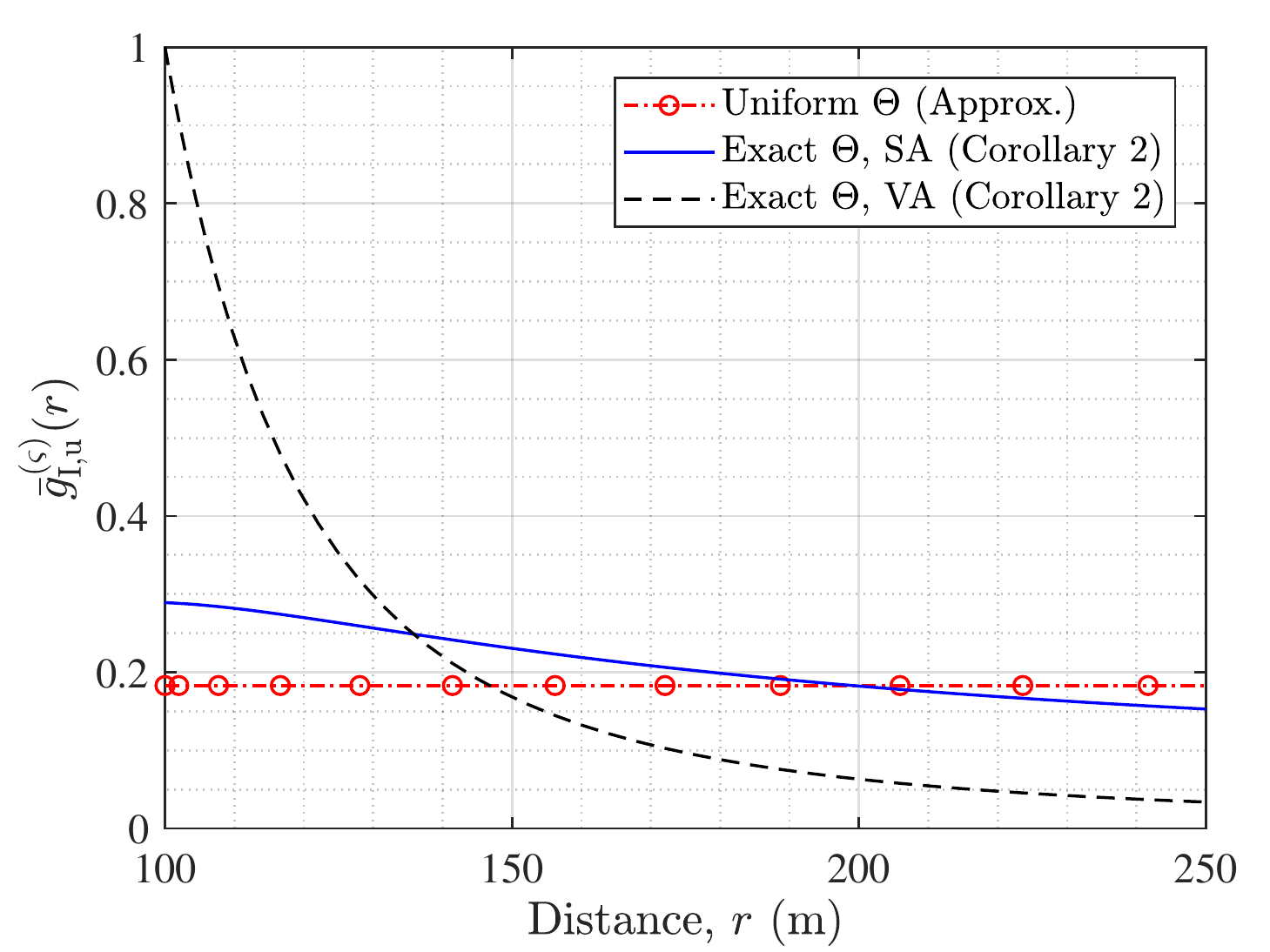}\label{fig:3b}}
    \caption{Impacts of uniform OBA assumption on $f_{\Theta|L}(\theta)$ and $\bar{g}_{\mathrm{I},\mathrm{u}}^{\left( \varsigma \right)}\left( r \right)$ when $\lambda_{\mathrm{u}}=20/\text{km}^2$, $\vartheta_{3\mathrm{dB}}=60^\circ$, and $G_{\mathrm{M},\mathrm{u}}=1$.}
    \label{fig:3}
\end{figure}

\begin{remark}
To simplify the evaluation of UAV interference, one widely used assumption is $\Theta\sim\mathrm{Uniform}(0,\pi)$ \cite{YiLiuBodaNallKara19,LiuHoWu19,AzarGeraGarcPoll20,_ShiDeng21,CherAlzeYaniYong21}, i.e., $F_{\left. \Theta \right|L}\left( \theta \right) ={\theta}/{\pi}$, $f_{\left. \Theta \right|L}\left( \theta \right) ={1}/{\pi},\theta \in \left[ 0,\pi \right]$. 
This assumption is reasonable in terrestrial networks, where the altitudes of TBSs and users are comparable, and thus $\Theta$ is uniformly distributed. 
However, in UAV-assisted cellular networks, as shown in Fig.~\ref{fig:2}, the altitude disparity between UAVs and terrestrial users induces a three-dimensional structure of $\Theta$. 
This results in location-dependent $\Theta$, i.e., the distribution of $\Theta$ is closely related to the distance between the interfering UAV and the typical user. In this case, the assumption of uniform distributed $\Theta$ is no longer appropriate.

In Fig.~\ref{fig:3}, we compare $f_{\Theta|L}(\theta)$ and $\bar{g}_{\mathrm{I},\mathrm{u}}^{\left( \varsigma \right)}\left( r \right)$ of an interfering UAV with and without uniform OBA assumption. 
It is observed that when $H_{\mathrm{u}}=0$, $\Theta$ is uniformly distributed in $[0,\pi]$ and this assumption is accurate. 
With the increase of UAV altitude, the UAV association area shrinks and thus a UAV tends to point its beam vertically downward. 
This reduces the randomness of $\Theta$ and makes the uniform OBA assumption no longer appropriate. 
These observations validate the necessity of considering the realistic distribution of $\Theta$ in the stochastic geometry-based analysis framework.
\end{remark}

\section{SINR MD Analysis}\label{sec:meta_distribution}
In this section, we present the main theoretical results for UAV-assisted cellular networks. The SINR MD-based analytical framework is first described in Section~\ref{subsec:4A}. The expressions of CSP moments for both $\mathrm{SA}$ and $\mathrm{VA}$ scenarios are then derived in Section~\ref{subsec:4B}, based on which we analyze the asymptotic behaviors of association probability as well as CSP moments when $\lambda_{k}\rightarrow\infty$ in Section~\ref{subsec:4C}. Finally, we turn our attention to several relevant special cases in Section~\ref{subsec:4D} for the sake of simple and intuitive results.

\subsection{Analytical Framework}\label{subsec:4A}

From \eqref{eq:sinr_define} we can see that the randomness of SINR stems from the realizations of network geometry $\Phi _{\mathrm{b}}\cup \Phi _{\mathrm{u}}$ and small-scale fading $\mathsf{H}_X$. 
To conduct a fine-grained analysis of SINR, the basic idea is to deal with the randomness of BSs locations and small-scale fading in two steps. 
Conditioned on $\Phi _{\mathrm{b}}$ and $\Phi _{\mathrm{u}}$, the CSP is defined as
\begin{align}
    \mathcal{P} _{\mathrm{s}}\left( \gamma |\Phi _{\mathrm{b}},\Phi _{\mathrm{u}} \right) =\mathbb{P} \left( \mathrm{SINR}>\gamma |\Phi _{\mathrm{b}},\Phi _{\mathrm{u}} \right),\label{eq:define_csp}
\end{align}
where $\gamma$ is the SINR threshold. 
Here $\mathcal{P} _{\mathrm{s}}\left( \gamma |\Phi _{\mathrm{b}},\Phi _{\mathrm{u}} \right)$ characterizes the CCDF of SINR given the snapshot of network geometry and can be viewed as a random variable with respect to $\Phi _{\mathrm{b}}$ and $\Phi _{\mathrm{u}}$. 
Then, the SINR MD describes the CDF of $\mathcal{P} _{\mathrm{s}}\left( \gamma |\Phi _{\mathrm{b}},\Phi _{\mathrm{u}} \right)$, i.e., 
\begin{align}
    \bar{F}_{\mathcal{P} _{\mathrm{s}}\left( \gamma \right)}\left( x \right) =\mathbb{P} \left( \mathcal{P} _{\mathrm{s}}\left( \gamma |\Phi _{\mathrm{b}},\Phi _{\mathrm{u}} \right) \geq x \right).\label{eq:define_md}
\end{align}
Due to the ergodicity of the point process, $\bar{F}_{\mathcal{P} _{\mathrm{s}}\left( \gamma \right)}\left( x \right)$ can be physically interpreted as the fraction of active links with the CSP of threshold $\gamma$ greater than $x$.

To calculate the SINR MD, one may resort to the Gil-Pelaez inversion theorem \cite{GilP51}, i.e.,
\begin{align}
    \bar{F}_{\mathcal{P} _{\mathrm{s}}\left( \gamma \right)}\left( x \right) =\frac{1}{2}+\frac{1}{\pi}\int_0^{\infty}{\frac{\mathrm{Im}\left[ e^{-\mathrm{i}t\log x}M _{\mathrm{i}t} \right]}{t}\mathrm{d}t},\quad x\in [0,1],\label{eq:gil-pelaez}
\end{align}
where $M_b$ is the $b$-th moment of $\mathcal{P} _{\mathrm{s}}\left( \gamma \right)$, $\mathrm{i} =\sqrt{-1}$ is the imaginary unit, and $\mathrm{Im}[\cdot]$ denotes the imaginary part of a complex number.
Although \eqref{eq:gil-pelaez} provides an exact expression of $\bar{F}_{\mathcal{P} _{\mathrm{s}}\left( \gamma \right)}\left( x \right)$, it involves tedious numerical calculations of the imaginary moments and integrations \cite{YuCuiWangLiTao20}. 
In search of a simpler alternative method, we approximate the SINR MD with standard beta distribution by matching their first two moments \cite{Haen16,DengHaen17}, which yields 
\begin{align}
    \bar{F}_{\mathcal{P} _{\mathrm{s}}\left( \gamma \right)}\left( x \right)\approx 1-I_x\left( \frac{M _1\left( M _1-M _2 \right)}{M _2-M _{1}^{2}},\frac{\left( 1-M _1 \right) \left( M _1-M _2 \right)}{M _2-M _{1}^{2}} \right),\quad x\in [0,1],\label{eq:beta_approx}
\end{align}
where $I_x(\cdot,\cdot)$ is the regularized incomplete beta function.

\subsection{General Results}\label{subsec:4B}

We first focus on the derivation of CSP's $b$-th moment. 
According to \eqref{eq:define_csp} and the law of total probability, $\mathcal{P} _{\mathrm{s}}\left( \gamma |\Phi _{\mathrm{b}},\Phi _{\mathrm{u}} \right)$ is formulated by
\begin{align}
    \mathcal{P} _{\mathrm{s}}\left( \gamma |\Phi _{\mathrm{b}},\Phi _{\mathrm{u}} \right) =\sum_{k\in \mathcal{K}}{A_{k}^{\left( \varsigma \right)}\mathcal{P} _{\mathrm{s}|k}\left( \gamma |\Phi _{\mathrm{b}},\Phi _{\mathrm{u}} \right)},\label{eq:csp-m1}
\end{align}
where $\mathcal{P} _{\mathrm{s}|k}\left( \gamma |\Phi _{\mathrm{b}},\Phi _{\mathrm{u}} \right)$ is the CSP conditioned on the typical user being associated with the $k$-th tier. 
Specifically, $\mathcal{P} _{\mathrm{s}|k}\left( \gamma |\Phi _{\mathrm{b}},\Phi _{\mathrm{u}} \right)$ is evaluated as
\begin{align}
    \mathcal{P} _{\mathrm{s}|k}\left( \gamma |\Phi _{\mathrm{b}},\Phi _{\mathrm{u}} \right) &\coloneqq\mathbb{P} \left( \mathrm{SINR}>\gamma |\Phi _{\mathrm{b}},\Phi _{\mathrm{u}},X^{\star}\in\Phi_k \right) \notag
    \\
    &=\mathbb{P} \left( \left. \mathsf{H}_{X^{\star}}>\frac{\gamma \left( N_0+\sum_{\ell \in \mathcal{K}}{I_{\ell}} \right)}{l_{k}^{\left( \varsigma \right)}\left( \left\| X^{\star} \right\| \right)} \right|\Phi _{\mathrm{b}},\Phi _{\mathrm{u}},X^{\star}\in \Phi _k \right) \notag
    \\
    &\overset{(a)}{\le}1-\mathbb{E} _{\mathsf{H},G}\left[ \left[ 1-\exp \left( -\frac{\gamma \phi _k\left( N_0+\sum_{\ell \in \mathcal{K}}{I_{\ell}} \right)}{l_{k}^{\left( \varsigma \right)}\left( Y_{0,k} \right)} \right) \right] ^{\mathsf{M}_k} \right],\label{eq:csp_k_expression}
\end{align}
{where $\phi _k\coloneqq\mathsf{M} _k\left( \mathsf{M} _k! \right) ^{-1/\mathsf{M} _k}$ and (a) follows from the tight upper bound of the CDF of a normalized gamma random variable \cite[Theo.~1]{Alze97}.} 
It is noticeable that the equality in (a) holds for $\mathsf{M}_k=1$, which is achieved when the typical user is associated with a TBS, i.e., $k=\mathrm{b}$.

The $b$-th moment of $\mathcal{P} _{\mathrm{s}}\left( \gamma |\Phi _{\mathrm{b}},\Phi _{\mathrm{u}} \right)$ for an individual link is consequently formulated by
\begin{align}
    M_{b}^{\left( \varsigma \right)}\left( \gamma \right) =\sum_{k\in \mathcal{K}}{A_{k}^{\left( \varsigma \right)}M_{b|k}^{\left( \varsigma \right)}\left( \gamma \right)},\quad \varsigma \in \{\mathrm{SA},\mathrm{VA}\}.\label{eq:mbk_to_mb}
\end{align}
The subsequent theorem presents the explicit expressions of $M_b^{(\varsigma)}\left( \gamma \right)$ for $\mathrm{SA}$ and $\mathrm{VA}$ scenarios.

\begin{theorem}[Moments for $\mathrm{SA}$ and $\mathrm{VA}$ Scenarios]\label{theo:mbk_sa_va}
{In UAV-assisted cellular networks, when UAVs are equipped with antennas of type $\varsigma\in\{\mathrm{SA},\mathrm{VA}\}$, for an active typical link, the $b$-th moment of $\mathcal{P} _{\mathrm{s}}\left( \gamma |\Phi _{\mathrm{b}},\Phi _{\mathrm{u}} \right)$ is approximated by 
\begin{align}
    M_{b}^{\left( \varsigma \right)}\left( \gamma \right) &\approx \sum_{k\in \mathcal{K}}{\sum_{n=0}^b{\sum_{m=0}^{\mathsf{M}_kn}{\binom{b}{n}}}\binom{\mathsf{M}_kn}{m}\left( -1 \right) ^{n+m}\int_0^{\infty}{\bar{\lambda}_k\left( y \right)}} \notag
    \\
    &\quad\times \exp \left\{ -\frac{m\gamma \phi _kN_0}{l_{k}^{\left( \varsigma \right)}\left( y \right)}-\sum_{\ell \in \mathcal{K}}{\left[ \bar{\Lambda}_{\ell}\left( \left[ 0,\chi _{k,\ell}^{\left( \varsigma \right)}\left( y \right) \right] \right) +\mathcal{U}\left( \gamma ,y \right) \right]} \right\} \,\mathrm{d}y, \label{eq:mb_general}
\end{align}
where 
\begin{align}
    \mathcal{U} \left( \gamma ,y \right) =\int_{\chi _{k,\ell}^{\left( \varsigma \right)}\left( y \right)}^{\infty}{\left[ 1-\mathcal{M} _{\mathsf{H}_{\ell}}\left( m\gamma \phi _kh_{\ell}^{\left( \varsigma \right)}\left( r \right) /l_{k}^{\left( \varsigma \right)}\left( y \right) \right) \right] \bar{\lambda}_{\ell}\left( r \right) \,\mathrm{d}r} \label{eq:u_mk_sa_va}
\end{align}
and $\mathcal{M} _{\mathsf{H}_{\ell}}\left( s \right) =\left( 1+s/\mathsf{M}_{\ell} \right) ^{-\mathsf{M}_{\ell}}$. 
Here $l_{k}^{\left( \varsigma \right)}\left( \cdot \right)$ and $h_{k}^{\left( \varsigma \right)}\left( \cdot \right)$ are given in \eqref{eq:l_k} and \eqref{eq:h_k}, respectively. 
}
\end{theorem}
\begin{IEEEproof}
See Appendix~\ref{app:theo_mbk_sa_va}.
\end{IEEEproof}

The SINR MD is then obtained by substituting \eqref{eq:mb_general} into \eqref{eq:beta_approx}. 

{\begin{remark}[Extension to General Propagation Scenario]
The analytical framework in \textbf{Theorem~\ref{theo:mbk_sa_va}} can be extended to the general propagation scenario of interfering links. 
Specifically, for each $k\in\mathcal{K}$, \textit{1)} consider arbitrary bounded, non-negative, and non-increasing deterministic path-loss function $L_k(r)$; 
\textit{2)} consider arbitrary normalized random small-scale fading gain $\mathsf{H}_k$ with moment generating function $\mathcal{M} _{\mathsf{H}_{k}}\left( s \right)$. 
Then, substituting \eqref{eq:l_k}, \eqref{eq:h_k}, and $\mathcal{M} _{\mathsf{H}_{k}}\left( s \right)$ into \eqref{eq:mb_general} yields the final result. 
\end{remark}}

The mean local delay $D\left( \gamma \right)$, defined as the average number of transmission attempts needed until the first successful transmission occurs \cite{ZhonZhanHaen14}, is another important metric to evaluate refined link performance. Assuming each transmission attempt is independent, according to the mean of geometric distribution, $D\left( \gamma \right)$ is formulated as
\begin{align}
    D^{\left( \varsigma \right)}\left( \gamma \right) =\sum_{k\in \mathcal{K}}{A_{k}^{\left( \varsigma \right)}D_{k}^{\left( \varsigma \right)}\left( \gamma \right)}=\sum_{k\in \mathcal{K}}{A_{k}^{\left( \varsigma \right)}\mathbb{E} \left[ \frac{1}{\mathcal{P} _{\mathrm{s}|k}\left( \gamma |\Phi _{\mathrm{b}},\Phi _{\mathrm{u}} \right)} \right]}=\sum_{k\in \mathcal{K}}{A_{k}^{\left( \varsigma \right)}M_{-1|k}^{\left( \varsigma \right)}\left( \gamma \right)},
\end{align}
which can be directly obtained based on the results of association probability and CSP moments.

\subsection{Asymptotic Behavior}\label{subsec:4C}

The recent advances in drone manufacturing make it possible to deploy widely UAVs for wireless communication purposes \cite{WuXuZengNgAlDh21}. 
When the network density goes to infinity, in this subsection, we analyze the asymptotic behavior of UAV-assisted cellular networks in terms of association probability and CSP moments.

{\begin{corollary}[Asymptotic Behavior when $\lambda_{\mathrm{b}}\rightarrow\infty$]\label{coro:asy_inf_lambdab}
As $\lambda_{\mathrm{b}}\rightarrow\infty$, it holds that 
\begin{align}
    \lim_{\lambda _{\mathrm{b}}\rightarrow \infty} A_{\mathrm{b}}^{\left( \varsigma \right)}=\exp \left\{ -2\pi \lambda _{\mathrm{u}}\left[ \int_{H_{\mathrm{u}}}^{\chi _{\mathrm{b},\mathrm{L}}^{\left( \varsigma \right)}\left( H_{\mathrm{b}} \right)}{p_{\mathrm{L}}\left( x \right) x\,\mathrm{d}x}+\int_{H_{\mathrm{u}}}^{\chi _{\mathrm{b},\mathrm{N}}^{\left( \varsigma \right)}\left( H_{\mathrm{b}} \right)}{p_{\mathrm{N}}\left( x \right) x\,\mathrm{d}x} \right] \right\} \label{eq:A_b_asym1}
\end{align}
and $\lim_{\lambda _{\mathrm{b}}\rightarrow \infty} M_{b|k}^{\left( \varsigma \right)}\left( \gamma \right) =0$, $\forall b\in \mathbb{N} ^+$, $k\in\mathcal{K}$. 
Specifically, when $P_{\mathrm{b}}G_{\mathrm{M},\mathrm{b}}\kappa _{\mathrm{b}}H_{\mathrm{b}}^{-\alpha _{\mathrm{b}}}\ge P_{\mathrm{u}}G_{\mathrm{M},\mathrm{u}}\kappa _{\mathrm{L}}H_{\mathrm{u}}^{-\alpha _{\mathrm{L}}}$, $\lim_{\lambda _{\mathrm{b}}\rightarrow \infty} A_{\mathrm{b}}^{\left( \varsigma \right)}=1$. 
\end{corollary}
\begin{IEEEproof}
When $\lambda_{\mathrm{b}}\rightarrow\infty$, the horizontal distance from the typical user to the nearest TBS tends to zero and thus $R_{0,\mathrm{b}}=H_{\mathrm{b}}$. 
In this case, according to \textbf{Theorem~\ref{theo:asso_prob}}, we have 
\begin{align}
    \lim_{\lambda _{\mathrm{b}}\rightarrow \infty} A_{\mathrm{b}}^{\left( \varsigma \right)} =\mathbb{E} _{R_{0,\mathrm{b}}}\left[ \prod_{\ell \in \left\{ \mathrm{L},\mathrm{N} \right\}}{\bar{F}_{R_{0,\ell}}\left( \chi _{\mathrm{b},\ell}^{\left( \varsigma \right)}\left( R_{0,\mathrm{b}} \right) \right)} \right] =\prod_{\ell \in \left\{ \mathrm{L},\mathrm{N} \right\}}{\bar{F}_{R_{0,\ell}}\left( \chi _{\mathrm{b},\ell}^{\left( \varsigma \right)}\left( H_{\mathrm{b}} \right) \right)}. \label{eq:A_b_asy1_p1}
\end{align}
Substituting \eqref{eq:r0k_cdf} into \eqref{eq:A_b_asy1_p1} yields \eqref{eq:A_b_asym1}. 
{Recalling the proof in Appendix~\ref{app:theo_mbk_sa_va}, since $\mathcal{M} _{\mathsf{H}_{\mathrm{b}}}\left( s \right) \in \left[ 0,1 \right]$, $\forall s\in \left[ 0,\infty \right)$, we have $\lim_{\lambda _{\mathrm{b}}\rightarrow \infty} \mathcal{L} _{I_{\mathrm{b}}}\left( s \right) =0$. 
For any $k\in \mathcal{K}$, $m\in \mathbb{N} _{0}^{+}$, and $b\in\mathbb{N}^{+}$, we further conclude from \eqref{eq:proof_mb_2} and \eqref{eq:proof_mb_3} that $\lim_{\lambda _{\mathrm{b}}\rightarrow \infty} \mathcal{T} _{k,m}\left( \gamma \right) =0$ and $\lim_{\lambda _{\mathrm{b}}\rightarrow \infty} M_{b|k}^{\left( \varsigma \right)}\left( \gamma \right) =0$.}

For each $\ell \in \left\{ \mathrm{L},\mathrm{N} \right\}$, according to the definition of $\chi _{\mathrm{b},\ell}^{\left( \varsigma \right)}\left( r \right)$ and its monotonicity, $\chi _{\mathrm{b},\ell}^{\left( \varsigma \right)}\left( H_{\mathrm{b}} \right)$ is bounded by $H_{\mathrm{u}}\le \chi _{\mathrm{b},\ell}^{\left( \varsigma \right)}\left( H_{\mathrm{b}} \right) \le \chi _{\mathrm{b},\mathrm{L}}^{\left( \varsigma \right)}\left( H_{\mathrm{b}} \right)$. 
When $P_{\mathrm{b}}G_{\mathrm{M},\mathrm{b}}\kappa _{\mathrm{b}}H_{\mathrm{b}}^{-\alpha _{\mathrm{b}}}\ge P_{\mathrm{u}}G_{\mathrm{M},\mathrm{u}}\kappa _{\mathrm{L}}H_{\mathrm{u}}^{-\alpha _{\mathrm{L}}}$, we further have $\chi _{\mathrm{b},\mathrm{L}}^{\left( \varsigma \right)}\left( H_{\mathrm{b}} \right) \le H_{\mathrm{u}}$, which concludes $\chi _{\mathrm{b},\ell}^{\left( \varsigma \right)}\left( H_{\mathrm{b}} \right) =H_{\mathrm{u}}$ and $\lim_{\lambda _{\mathrm{b}}\rightarrow \infty} A_{\mathrm{b}}^{\left( \varsigma \right)} =1$. 
\end{IEEEproof}

\begin{corollary}[Asymptotic Behavior when $\lambda_{\mathrm{u}}\rightarrow\infty$]\label{coro:asy_inf_lambdau}
As $\lambda_{\mathrm{u}}\rightarrow\infty$, it holds that 
$\lim_{\lambda _{\mathrm{u}}\rightarrow \infty} A_{\mathrm{b}}^{\left( \varsigma \right)}=1-e^{-\pi \lambda _{\mathrm{b}}[ ( \chi _{\mathrm{L},\mathrm{b}}^{\left( \varsigma \right)}\left( H_{\mathrm{u}} \right) ) ^2-H_{\mathrm{b}}^{2} ]}$, 
$\lim_{\lambda _{\mathrm{u}}\rightarrow \infty} A_{\mathrm{L}}^{\left( \varsigma \right)}=e^{-\pi \lambda _{\mathrm{b}}[ ( \chi _{\mathrm{L},\mathrm{b}}^{\left( \varsigma \right)}\left( H_{\mathrm{u}} \right) ) ^2-H_{\mathrm{b}}^{2} ]}$, 
$\lim_{\lambda _{\mathrm{u}}\rightarrow \infty} A_{\mathrm{N}}^{\left( \varsigma \right)}=0$, and 
$\lim_{\lambda _{\mathrm{b}}\rightarrow \infty} M_{b|k}^{\left( \varsigma \right)}\left( \gamma \right) =0$, $\forall b\in \mathbb{N} ^+$, $k\in\mathcal{K}$. 
Specifically, when $P_{\mathrm{b}}G_{\mathrm{M},\mathrm{b}}\kappa _{\mathrm{b}}r^{-\alpha _{\mathrm{b}}}\le P_{\mathrm{u}}G_{\mathrm{M},\mathrm{u}}\kappa _{\mathrm{L}}H_{\mathrm{u}}^{-\alpha _{\mathrm{L}}}$, $\lim_{\lambda _{\mathrm{u}}\rightarrow \infty} A_{\mathrm{L}}^{\left( \varsigma \right)}=1$. 
\end{corollary}
\begin{IEEEproof}
Since the LoS conditions between different A2G links are independent, $\lambda_{\mathrm{u}}\rightarrow\infty$ guarantees the existence of a UAV hovering above the typical user in LoS condition. 
In this case, none of other NLoS UAVs could provide a stronger average received power to the typical user. 
The results are then obtained by following the similar lines as the proof of \textbf{Corollary~\ref{coro:asy_inf_lambdab}}. 
\end{IEEEproof}}

\subsection{Special Cases}\label{subsec:4D}

Although the general result in \textbf{Theorem~\ref{theo:mbk_sa_va}} involves a double integral with complicated form, simple and intuitive expressions of CSP moments are possible when considering specific link conditions or deployments. We now turn our attention to several relevant special cases.

\subsubsection{Noise-Limited}
When the proposed network is operated in millimeter-wave frequency and the densities of TBSs and UAVs are small, due to the large bandwidth and sparse interferers, the network can be regarded as noise-limited \cite{YiLiuDengNall20}. Here we present the CSP moments and SNR MD in this special case.

\begin{corollary}[Noise-Limited]\label{coro:case-noise-limit}
When the UAV-assisted cellular network is noise-limited, the $b$-th moment of $\mathcal{P} _{\mathrm{s}}\left( \gamma |\Phi _{\mathrm{b}},\Phi _{\mathrm{u}} \right)$ is given as
\begin{align}
    M_{b}^{\left( \varsigma \right)} =\sum_{k\in \mathcal{K}}{A_{k}^{\left( \varsigma \right)}\int_0^{\infty}{\left[ \frac{\Gamma \left( \mathsf{M}_k,\gamma N_0\mathsf{M}_k/l_{k}^{\left( \varsigma \right)}\left( y \right) \right)}{\Gamma \left( \mathsf{M}_k \right)} \right] ^bf_{Y_{0,k}^{\left( \varsigma \right)}}\left( y \right) \mathrm{d}y}}.
\end{align}
The SNR MD is evaluated as 
\begin{align}
    \bar{F}_{\mathcal{P} _{\mathrm{s}}\left( \gamma \right)}\left( x \right) &=\sum_{k\in \mathcal{K}}{A_{k}^{\left( \varsigma \right)}F_{Y_{0,k}}\left( \dot{l}_{k}^{\left( \varsigma \right)}\left( \frac{\gamma N_0}{\omega _{k,x}} \right) \right)},
\end{align}
where $\omega _{k,x}$ is the solution of the equation $\Gamma \left( \mathsf{M}_k,\mathsf{M}_k\omega _{k,x} \right) =x\Gamma \left( \mathsf{M}_k \right)$.
\end{corollary}
\begin{IEEEproof}
By definition, for the users associated with the $k$-th tier, the CSP is expressed as
\begin{align}
    \mathcal{P} _{\mathrm{s}|k}\left( \gamma |\Phi _{\mathrm{b}},\Phi _{\mathrm{u}} \right) &=\mathbb{P} \left( \left. \mathsf{H}_{X^{\star}}>\frac{\gamma N_0}{l_{k}^{\left( \varsigma \right)}\left( Y_{0,k} \right)} \right|\Phi _{\mathrm{b}},\Phi _{\mathrm{u}},X^{\star}\in \Phi _k \right) \notag
    \\
    &\overset{(a)}{=}\frac{1}{\Gamma \left( \mathsf{M}_k \right)}\Gamma \left( \mathsf{M}_k,\frac{\gamma N_0\mathsf{M}_k}{l_{k}^{\left( \varsigma \right)}\left( Y_{0,k} \right)} \right),
\end{align}
where $\Gamma(\cdot,\cdot)$ is the upper incomplete gamma function and (a) is from the CDF of Gamma random variable. Then the SNR MD is formulated by
\begin{align}
    \bar{F}_{\mathcal{P} _{\mathrm{s}|k}\left( \gamma \right)}\left( x \right) &=\mathbb{P} \left( \mathcal{P} _{\mathrm{s}|k}\left( \gamma \right) >x \right) \notag
    \\
    &=\mathbb{P} \left( \frac{\gamma N_0}{l_{k}^{\left( \varsigma \right)}\left( Y_{0,k} \right)}<\omega _{k,x} \right) \notag
    \\
    &\overset{(b)}{=}\mathbb{P} \left( Y_{0,k}<\dot{l}_{k}^{\left( \varsigma \right)}\left( \frac{\gamma N_0}{\omega _{k,x}} \right) \right) \notag
    \\
    &=F_{Y_{0,k}}\left( \dot{l}_{k}^{\left( \varsigma \right)}\left( \frac{\gamma N_0}{\omega _{k,x}} \right) \right),
\end{align}
where (b) is due to the monotonicity of $l_{k}^{\left( \varsigma \right)}\left( r \right)$. The proof is completed by using the law of total probability.
\end{IEEEproof}

\begin{remark}
By using Alzer's inequality \cite{Alze97}, $\omega _{k,x}$ can be tightly upper bounded by $\omega _{k,x}\le -\frac{\ln ( 1-\left( 1-x \right) ^{1/\mathsf{M}_k} )}{\mathsf{M}_k\left( \mathsf{M}_k! \right) ^{-1/\mathsf{M}_k}}$.
\end{remark}

\subsubsection{Canonical Isotropic Antenna Pattern}
Although our main focus is on antenna patterns of type $\mathrm{SA}$ and $\mathrm{VA}$, as a baseline for comparison, we also study the case in which UAV antenna has the same radiation pattern in all directions \cite{BanaDhil22}. The CSP moments for this special case are given as follows.

\begin{corollary}[Canonical Isotropic Antenna Pattern]\label{coro:isotropic}
{When TBSs and UAVs are equipped with canonical isotropic antennas, the $b$-th moment of $\mathcal{P} _{\mathrm{s}}\left( \gamma |\Phi _{\mathrm{b}},\Phi _{\mathrm{u}} \right)$ is approximated by
\begin{align}
    M_{b}\left( \gamma \right) &\approx \sum_{k\in \mathcal{K}}{\sum_{n=0}^b{\sum_{m=0}^{\mathsf{M}_kn}{\binom{b}{n}}}\binom{\mathsf{M}_kn}{m}\left( -1 \right) ^{n+m}\int_0^{\infty}{\bar{\lambda}_k\left( y \right)}} \times \exp \Bigg\{ -\frac{m\gamma \phi _kN_0}{P_kG_{\mathrm{M},k}y^{-\alpha _k}} \notag
    \\ 
    &\quad\left. -\sum_{\ell \in \mathcal{K}}{\left[ \bar{\Lambda}_{\ell}\left( \left[ 0,\chi _{k,\ell}\left( y \right) \right] \right) +\int_{\chi _{k,\ell}\left( y \right)}^{\infty}{\left[ 1-\Big( 1+\frac{m\gamma \phi _k\eta _{k,\ell}}{\mathsf{M}_{\ell}r^{\alpha _{\ell}}y^{-\alpha _k}} \Big) ^{-\mathsf{M}_{\ell}} \right] \bar{\lambda}_{\ell}\left( r \right) \,\mathrm{d}r} \right]} \right\} \mathrm{d}y, 
\end{align}
where $\eta _{k,\ell} =\frac{P_{\ell}G_{\mathrm{M},\ell}\kappa _{\ell}}{P_kG_{\mathrm{M},k}\kappa _k}$ and $\chi _{k,\ell}\left( y \right) =\eta _{k,\ell}^{1/\alpha _{\ell}}y^{\alpha _k/\alpha _{\ell}}$.} 
\end{corollary}
\begin{IEEEproof}
The result is directly obtained by substituting $\vartheta _{3\mathrm{dB},\mathrm{b}}=\vartheta _{3\mathrm{dB},\mathrm{u}}=\infty$ into \eqref{eq:mb_general}. The superscript ``$(\varsigma)$'' is omitted here since the results for $\mathrm{SA}$ and $\mathrm{VA}$ scenarios are the same in this context.
\end{IEEEproof}

\subsubsection{Rayleigh Fading}

The assumption of Rayleigh fading for both G2G and A2G links has been applied in several pioneering literature due to its tractability \cite{ChenZhan20,_DengChenWei21,AtzeArnaKoun18}. 
Considering this assumption in UAV-assisted cellular networks, the $b$-th moment of CSP is simplified in the following corollary.

\begin{corollary}[Rayleigh Fading]\label{coro:mbk_rayleigh}
When the UAVs are equipped with antennas of type $\varsigma\in\{\mathrm{SA},\mathrm{VA}\}$ and the small-scale fading is subject to Rayleigh, i.e., $\mathsf{M}_k=1$, $k\in \mathcal{K}$, the $b$-th moment of $\mathcal{P} _{\mathrm{s}}\left( \gamma |\Phi _{\mathrm{b}},\Phi _{\mathrm{u}} \right)$ is given as 
{
\begin{align}
    M_{b}^{\left( \varsigma \right)}\left( \gamma \right) &=\sum_{k\in \mathcal{K}} \int_0^{\infty} \bar{\lambda}_k\left( y \right) \exp \left\{ -\frac{b\gamma N_0}{l_{k}^{\left( \varsigma \right)}\left( y \right)}-\sum_{\ell \in \mathcal{K}} \bigg[ \bar{\Lambda}_{\ell}\left( \left[ 0,\chi _{k,\ell}^{\left( \varsigma \right)}\left( y \right) \right] \right)\right.\notag 
    \\
    &\qquad\qquad\left.+\int_{\chi _{k,\ell}^{\left( \varsigma \right)}\left( y \right)}^{\infty}{\Big[ 1-\big( 1+\gamma h_{\ell}^{\left( \varsigma \right)}\left( r \right) /l_{k}^{\left( \varsigma \right)}\left( y \right) \big) ^{-b} \Big] \bar{\lambda}_{\ell}\left( r \right) \,\mathrm{d}r} \bigg] \right\} \,\mathrm{d}y, \label{eq:mb_iota_rayleigh}
\end{align}
where $l_{k}^{\left( \varsigma \right)}\left( \cdot \right)$ and $h_{k}^{\left( \varsigma \right)}\left( \cdot \right)$ are given in \eqref{eq:l_k} and \eqref{eq:h_k}, respectively.}
\end{corollary}
\begin{IEEEproof}
The proof follows the same lines as in \textbf{Theorem~\ref{theo:mbk_sa_va}} and is thus skipped here. 
\end{IEEEproof}

\begin{remark}
Since the small-scale fading gain $\mathsf{H}_X$ follows exponential distribution under Rayleigh fading assumption, the proof of \textbf{Corollary~\ref{coro:mbk_rayleigh}} involves neither Alzer's approximation nor Jensen's bound. Hence the expression of $M_{b}^{\left( \varsigma \right)}\left( \gamma \right)$ given in \eqref{eq:mb_iota_rayleigh} is accurate.
\end{remark}

\subsubsection{Primary User}
{Previous results are subject to the scenario that all users have the same priority. 
However, for the primary user that needs the best channel condition, it is better to assume that the serving UAV flies towards this primary user and then hovers above its head to transmit information \cite{YiLiuBodaNallKara19,YiLiuDengNall20,SenaDurrZhouYangDing20}. 
The following corollary presents the expression of $M_{b}^{\left( \varsigma \right)}\left( \gamma \right)$ in this situation. 
\begin{corollary}[Primary User]
In the situation that there is a UAV hovering above the typical user providing LoS service, the $b$-th moment of $\mathcal{P} _{\mathrm{s}}\left( \gamma |\Phi _{\mathrm{b}},\Phi _{\mathrm{u}} \right)$ has a similar expression to \eqref{eq:mb_general}, and the only difference is 
\begin{align}
    l_{k}^{\left( \varsigma \right)}\left( r \right) =
    \begin{cases}
        P_{\mathrm{b}}G_{\mathrm{b}}\left( \arccos \left( \frac{H_{\mathrm{b}}}{r} \right) \right) \kappa _{\mathrm{b}}r^{-\alpha _{\mathrm{b}}}, & k=\mathrm{b},\\
        P_{\mathrm{u}}G_{\mathrm{u}}\left( 0 \right) \kappa _kH_{\mathrm{u}}^{-\alpha _k}, & k\in \left\{ \mathrm{L},\mathrm{N} \right\} .\\
    \end{cases}
\end{align}
\end{corollary}
\begin{IEEEproof}
The proof follows the similar lines as \textbf{Theorem~\ref{theo:mbk_sa_va}} and is thus omitted here. 
\end{IEEEproof}
}

\section{Simulation and Discussions}\label{sec:simulation}

In this section, we provide simulation and numerical results to validate the accuracy of our theoretical analysis and to give design insights of UAV-assisted cellular networks. 
Specifically, we first validate the necessity of considering the exact distribution of OBA in the analysis. 
Next, the effects of UAV features and environment parameters on the system performance are investigated. Finally, different network deployments are compared to provide useful design insights. 
The simulations are performed in a circular shape simulation area of radius $2000$\,m. 
{Unless otherwise stated, we set the deployment parameters as $\lambda_{\mathrm{b}}=5/\text{km}^2$, $\lambda_{\mathrm{u}}=20/\text{km}^2$, $H_{\mathrm{b}}=20$\,m, and $H_{\mathrm{u}}=100$\,m. 
The power and antenna parameters are $P_{\mathrm{b}}=30$\,W, $P_{\mathrm{u}}=10$\,W, $N_0=10^{-8}$\,W, $G_{\mathrm{M},\mathrm{b}}=0$\,dB, $G_{\mathrm{M},\mathrm{u}}=0$\,dB, $\vartheta_{3\mathrm{dB},\mathrm{b}}=160^\circ$, $\vartheta_{3\mathrm{dB},\mathrm{u}}=60^\circ$, and $\mu_{\mathrm{SLA}}=20$\,dB. 
Besides, following \cite{KouzElSaDahrAlshAlNa21,AlHoKandLard14}, we consider urban environments with parameters $\mu _a =9.61$, $\mu _b =0.16$, $\kappa_{k}=1$, $\alpha_{\mathrm{b}}=3$, $\alpha_{\mathrm{L}}=2.5$, $\alpha_{\mathrm{N}}=4$, $\mathsf{M}_{\mathrm{b}}=1$, $\mathsf{M}_{\mathrm{L}}=3$, and $\mathsf{M}_{\mathrm{N}}=2$.}

\subsection{Validation}

\begin{figure}[!t]
    \centering
    \subfloat[SINR coverage probability]{\includegraphics[width=0.5\textwidth]{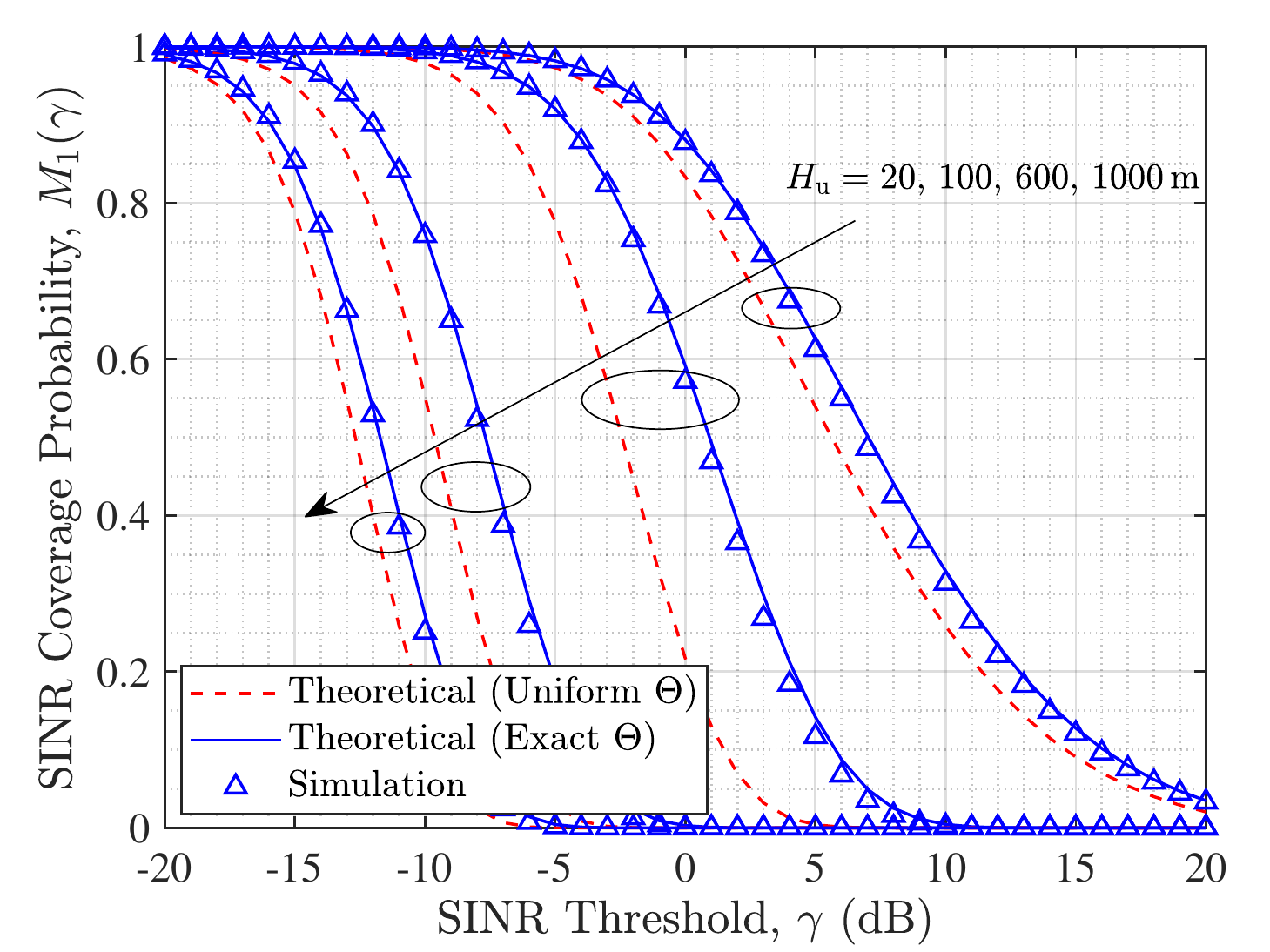}\label{fig:4a}}
    \hfil
    \subfloat[SINR MD]{\includegraphics[width=0.5\textwidth]{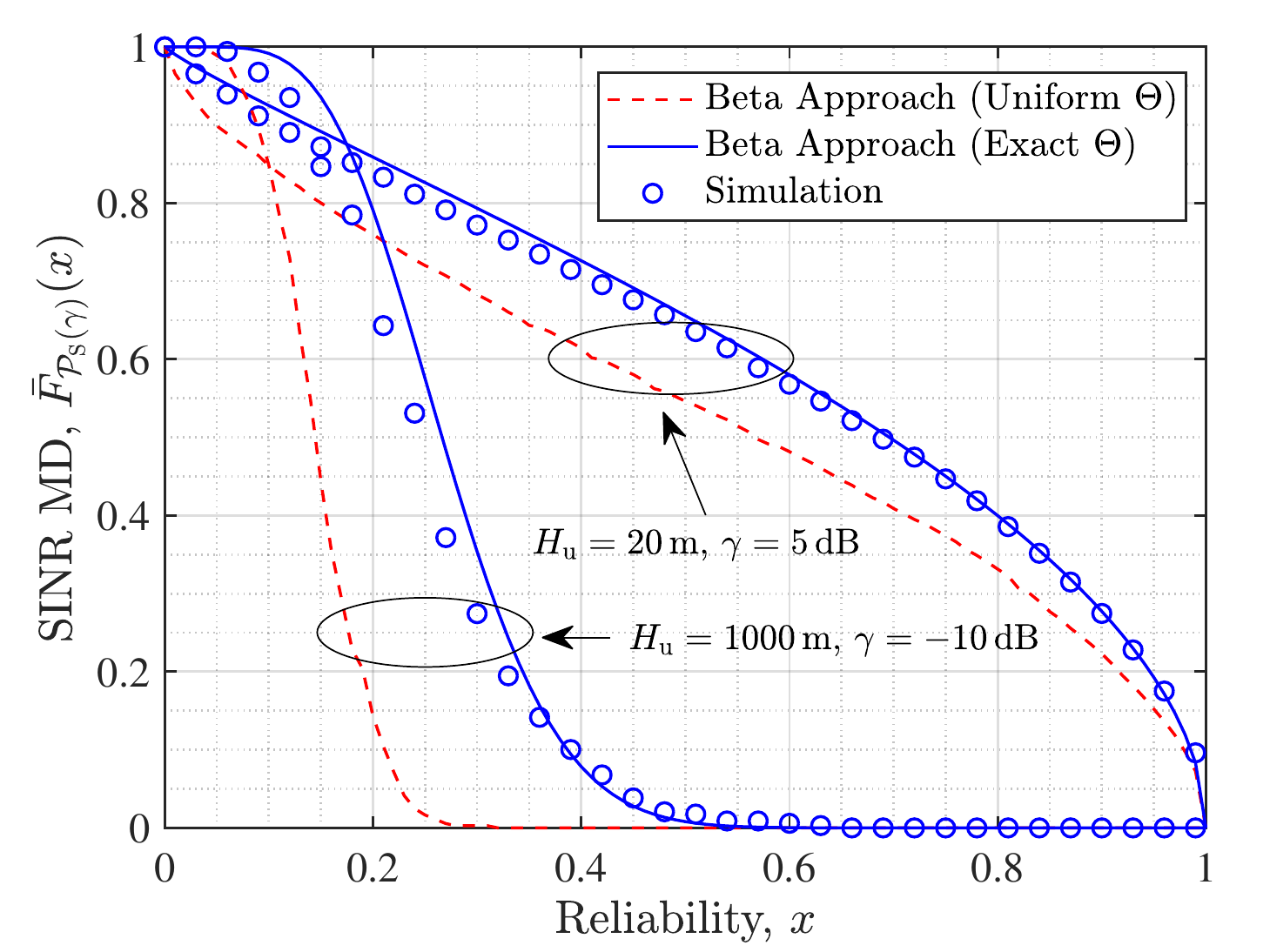}\label{fig:4b}}
    \caption{{Comparisons of the network performance under approximate and exact distribution of OBA when $\varsigma=\mathrm{SA}$, $\lambda_{\mathrm{b}}=5/\text{km}^2$, and $\lambda_{\mathrm{u}}=20/\text{km}^2$.}}
    \label{fig:4}
\end{figure}

In Fig.~\ref{fig:4}, we present the results of network performance under different approaches of OBA. 
{The close match between the simulation results and theoretical results with exact $\Theta$ validates the accuracy of the approximations in \eqref{eq:beta_approx} and \textbf{Theorem~\ref{theo:mbk_sa_va}}. 
Note that due to the incorporation of multi-tier interference, the standard beta approximation is less accurate in the regime of small reliability and SINR threshold. 
In this case, the approximations with higher degrees of freedom will help to close the gap \cite{DengHaen17,ShiGaoYangNiyaHan21}.} 
In general, we observe that adopting the exact OBA distribution leads to accurate SINR coverage probability $M_1(\gamma)$ and SINR MD $\bar{F}_{\mathcal{P}(\gamma)}(x)$ for aerial networks, while the uniform OBA assumption underestimates both metrics. 
This observation validates the necessity of accounting for the exact OBA distribution in the performance evaluation of UAV-assisted cellular networks. 
More specifically, Fig.~\ref{fig:4a} shows that when evaluating SINR coverage probability, the uniform OBA assumption is only relatively accurate in the regime of low and high UAV altitudes. 
The reason is that UAV beams are almost horizontally pointed for low altitude and the distances of serving link and dominant interfering links are not evidently distinguished for high altitude. 
In Fig.~\ref{fig:4b}, we further observe that even in this regime (e.g., $H_{\mathrm{u}}=20\,\text{m}$ or $H_{\mathrm{u}}=1000\,\text{m}$), there is an evident gap between the SINR MDs under uniform $\Theta$ assumption and exact $\Theta$ distribution.

\subsection{Impact of UAV Features}

\begin{figure}[!t]
    \centering
    \subfloat[Association probability]{\includegraphics[width=0.5\textwidth]{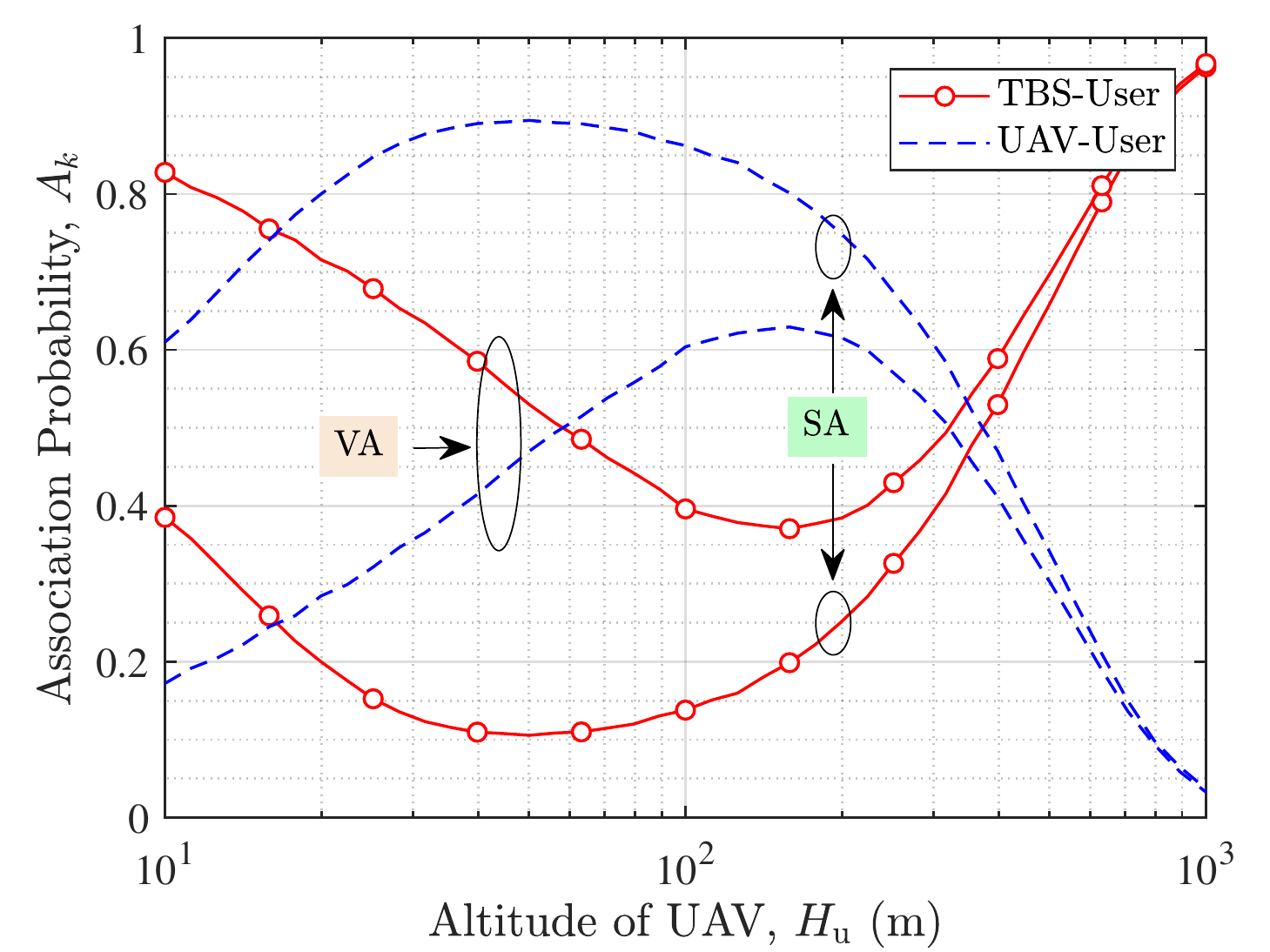}\label{fig:5a}}\hfil
    \subfloat[Mean and variance of CSP]{\includegraphics[width=0.5\textwidth]{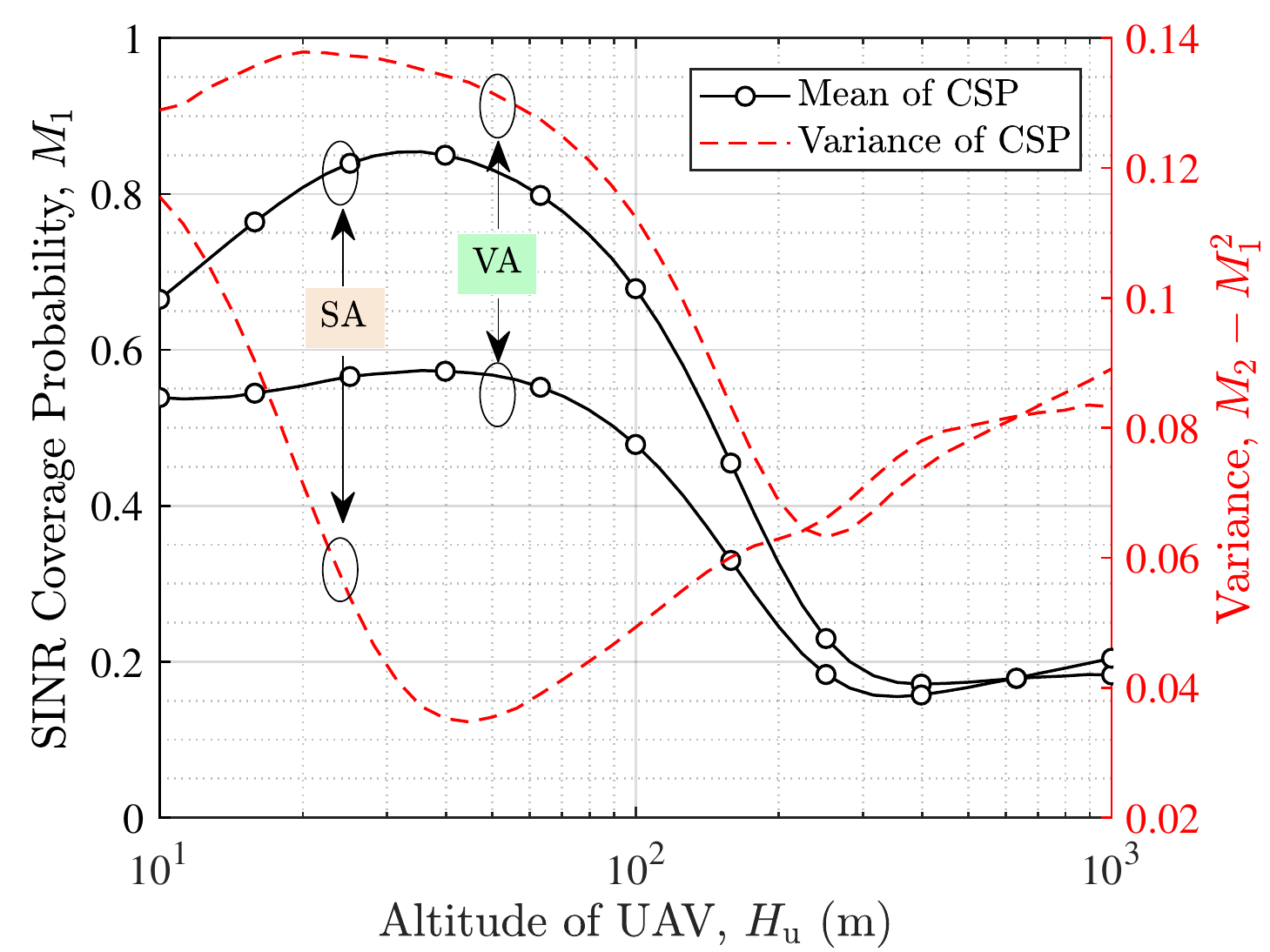}\label{fig:5b}}
    \caption{The impact of UAV altitude on the association probability, SINR coverage probability, and variance of CSP for both $\mathrm{SA}$ and $\mathrm{VA}$ scenarios when $\lambda_{\mathrm{b}}=5/\text{km}^2$ and $\lambda_{\mathrm{u}}=20/\text{km}^2$.}
    \label{fig:5}
\end{figure}

Fig.~\ref{fig:5} presents the network performance as a function of UAV altitude $H$. 
In general, for both $\mathrm{SA}$ and $\mathrm{VA}$ scenarios, when the UAV altitude increases from zero, the A2G links change from NLoS to LoS condition, which is beneficial to the association probability of UAVs and the coverage probability of the whole network. 
However, as the altitude increases further, the serving links from UAVs become longer and the interfering A2G links become more in LoS, which degrades the UAV association probability and the coverage probability. 
These two opposite factors are balanced at the optimal altitude to maximize the coverage performance, which is consistent with existing findings \cite{ZengZhan20BK,BanaDhil22}. 
Besides, Fig.~\ref{fig:5b} shows the results of CSP variance, which measures the fairness between individual links. 
The most interesting observation is that with the increase of UAV altitude, the CSP variance in the $\mathrm{VA}$ scenario has the same tendency as coverage probability, while the result in the $\mathrm{SA}$ scenario shows the opposite tendency. 
In other words, when UAVs are equipped with steerable antennas, by properly tuning the UAV altitude, it is expected to achieve the highest network coverage and the best user fairness at the same time.

\begin{figure}[!t]
    \centering
    \subfloat[Association probability]{\includegraphics[width=0.5\textwidth]{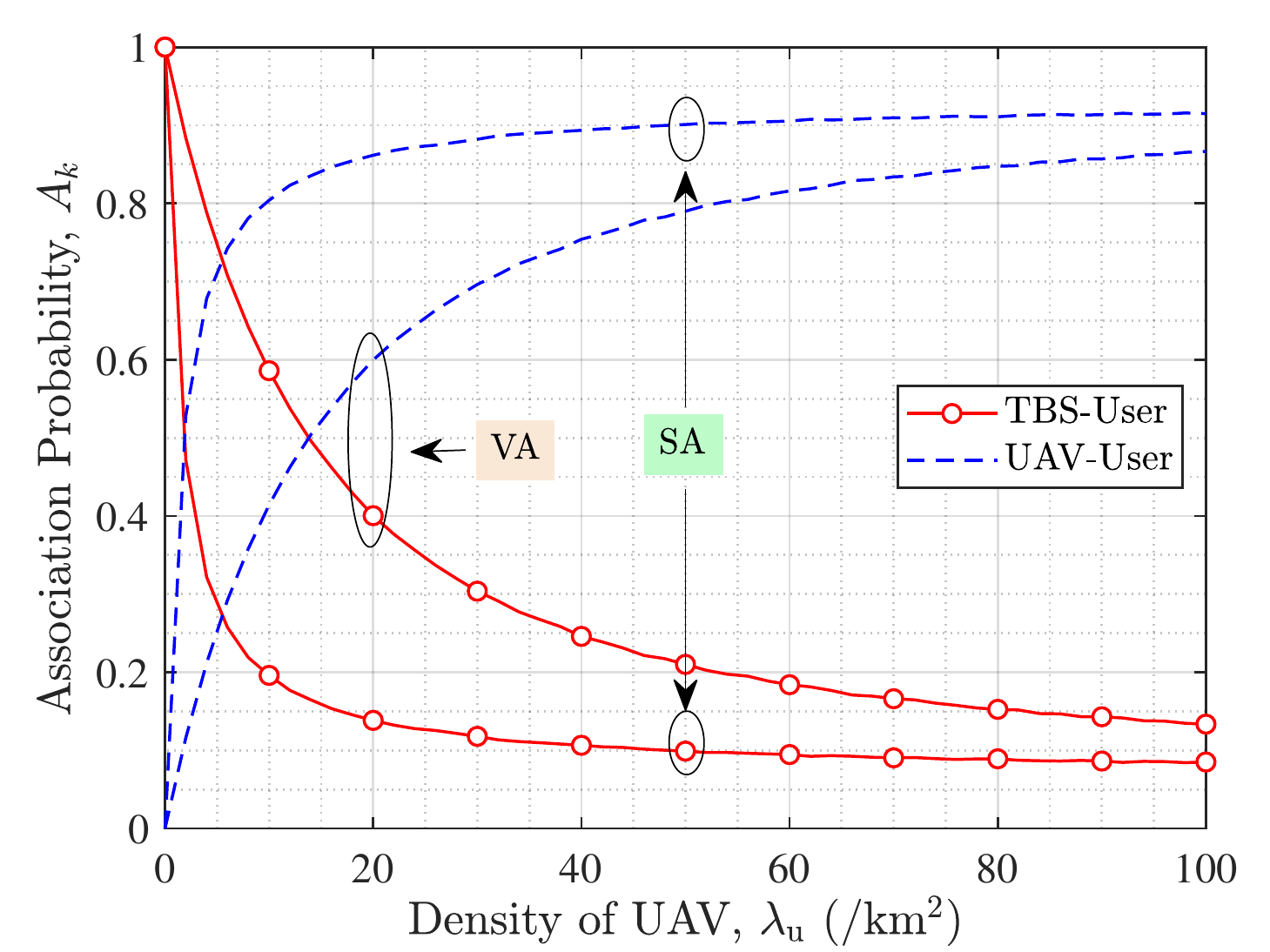}\label{fig:6a}}\hfil
    \subfloat[Mean and variance of CSP]{\includegraphics[width=0.5\textwidth]{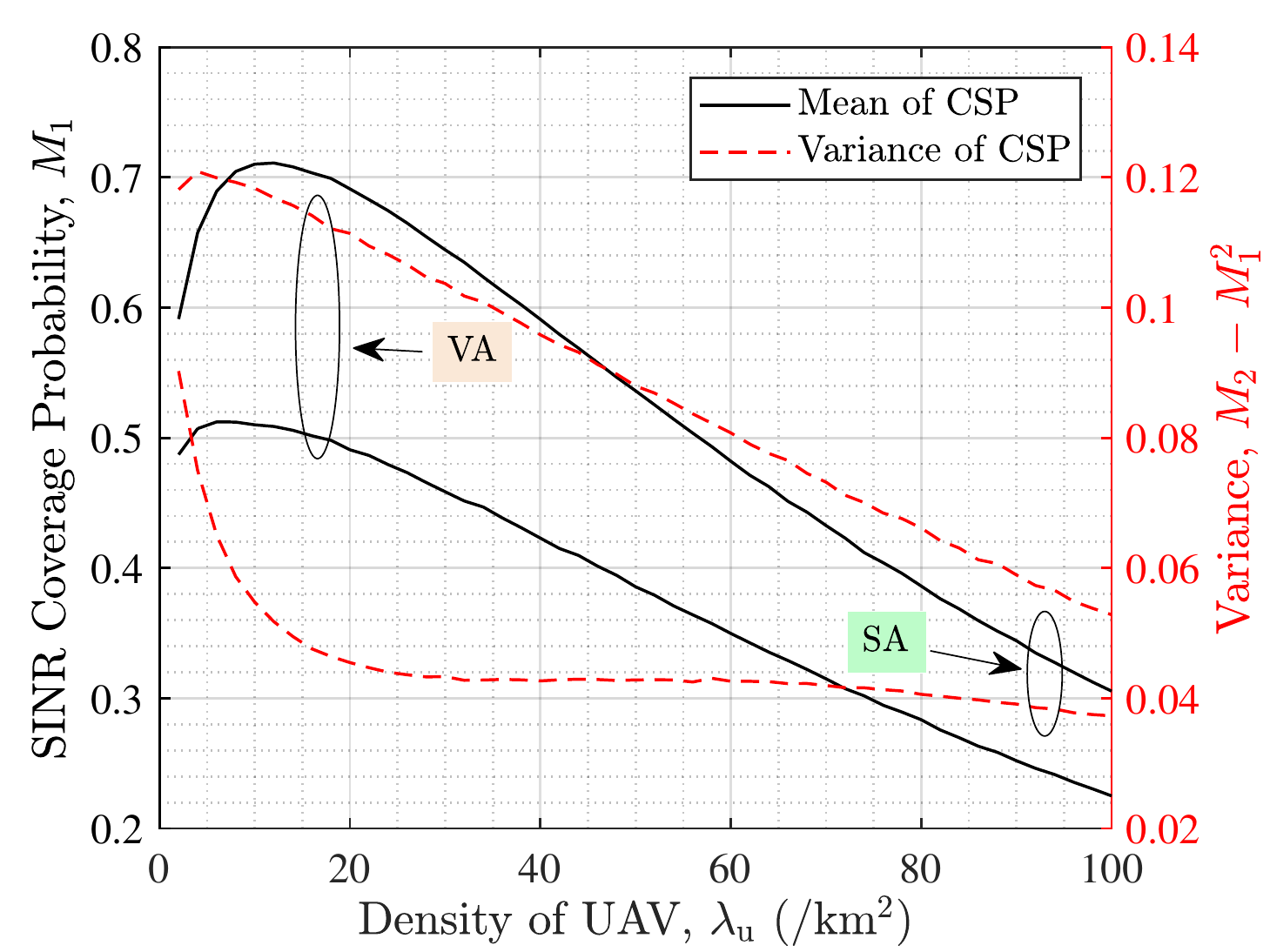}\label{fig:6b}}
    \caption{The impact of UAV density on the association probability, SINR coverage probability, and variance of CSP for both $\mathrm{SA}$ and $\mathrm{VA}$ scenarios when $\lambda_{\mathrm{b}}=5/\text{km}^2$ and $H_{\mathrm{u}}=100$\,\text{m}.}
    \label{fig:6}
\end{figure}

Fig.~\ref{fig:6} highlights the impact of UAV density on network performance. 
With more UAVs being deployed, the communication link to the nearest UAV becomes shorter and is more likely to be in LoS condition. Thus the association probability of UAVs monotonically increases with $\lambda_{\mathrm{u}}$. 
However, the increase of $\lambda_{\mathrm{u}}$ also brings more interfering UAVs, in both LoS and NLoS conditions. Thus the coverage probability is maximized at a particular UAV density and then starts decreasing. 
In fact, as stated in Section~\ref{subsec:4C}, the coverage probability will converge to zero when the network density goes to infinity in our framework. 
Moreover, we show that the CSP variance in $\mathrm{VA}$ scenario almost linearly decreases with $\lambda_{\mathrm{u}}$, while the CSP variance in $\mathrm{SA}$ scenario first rapidly decreases and then slowly converges to $0.04$ for high densities. 
Unlike Fig.~\ref{fig:5b}, the coverage probability and variance are not optimized simultaneously. Therefore, an appropriate design should establish a proper coverage and fairness trade-off.

\begin{figure}[!t]
    \centering
    \subfloat[Association probability]{\includegraphics[width=0.5\textwidth]{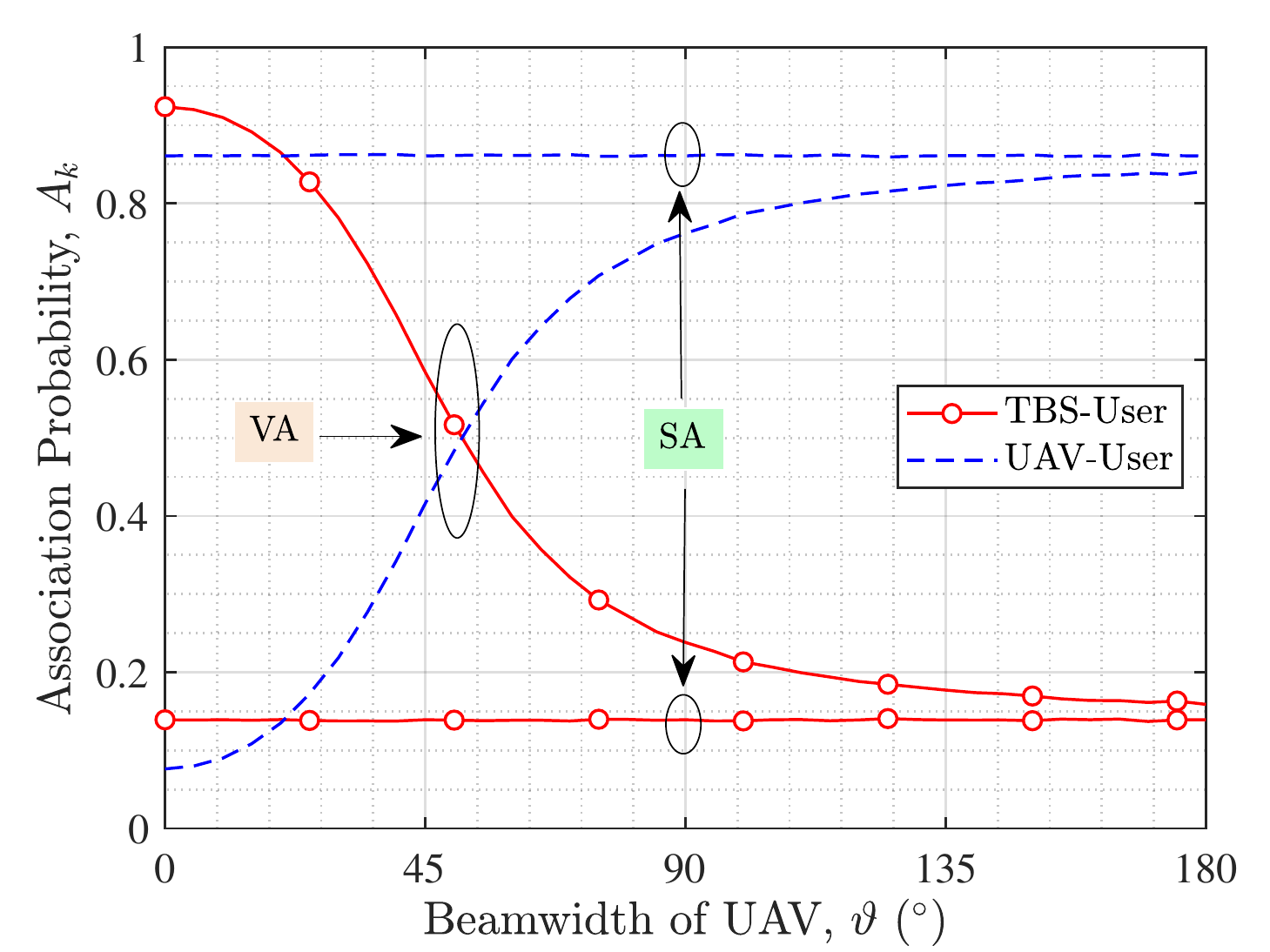}\label{fig:7a}}\hfil
    \subfloat[Mean and variance of CSP]{\includegraphics[width=0.5\textwidth]{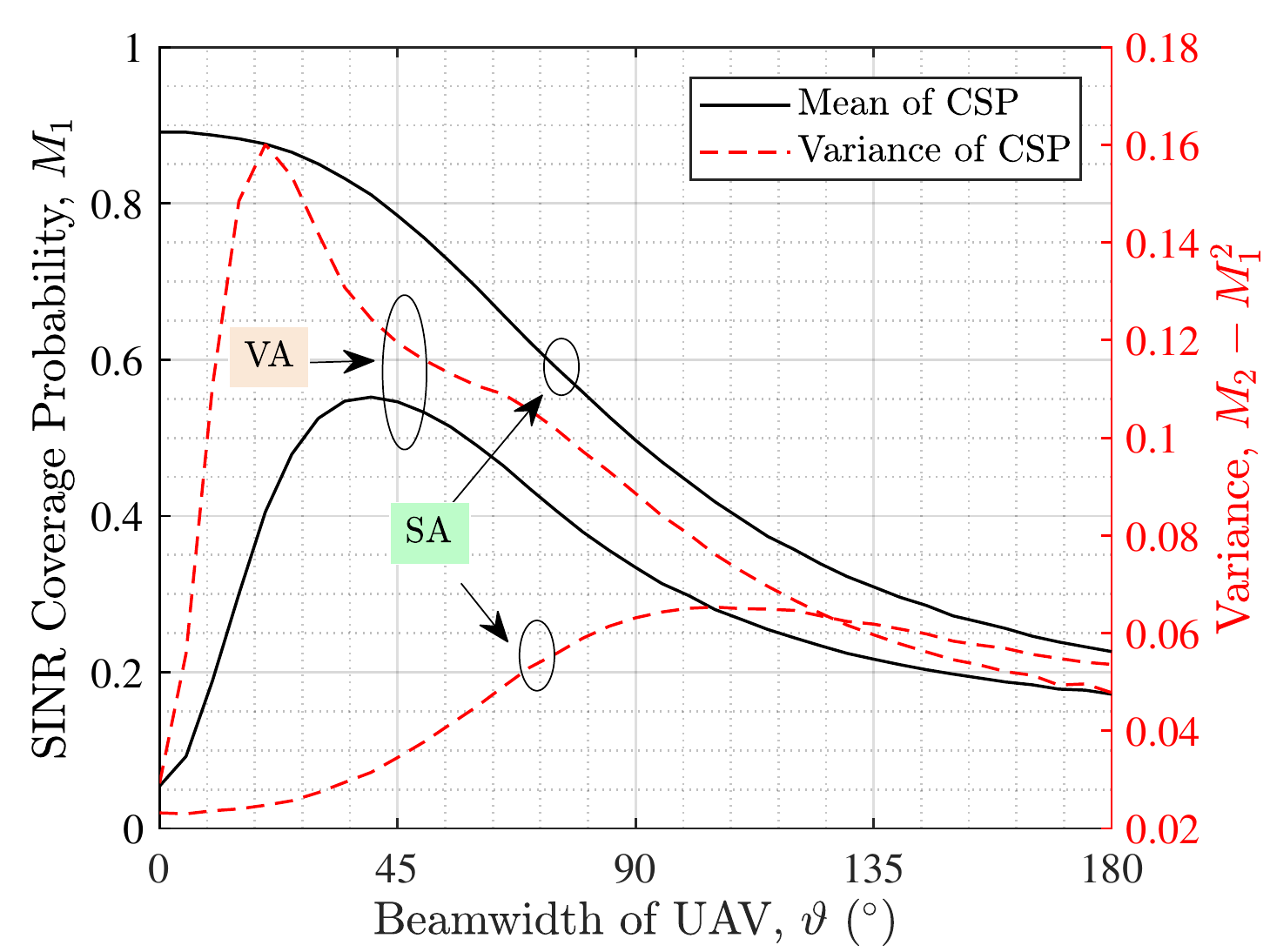}\label{fig:7b}}
    \caption{The impact of UAV beamwidth on the association probability, SINR coverage probability, and variance of CSP for both $\mathrm{SA}$ and $\mathrm{VA}$ scenarios when $\lambda_{\mathrm{b}}=5/\text{km}^2$, $\lambda_{\mathrm{u}}=20/\text{km}^2$, and $H_{\mathrm{u}}=100$\,\text{m}.}
    \label{fig:7}
\end{figure}

Fig.~\ref{fig:7} shows the impact of UAV beamwidth $\vartheta$ on the network performance. 
In the $\mathrm{VA}$ scenario, UAVs point their beams vertically downward. Thus with the increase of UAV beamwidth, there will be more candidate UAVs covering the typical user in their mainlobes, which is constructive to the association probability of UAVs and the coverage probability. 
However, further increasing UAV beamwidth results in stronger antenna gains from interfering UAVs, which is destructive to the coverage probability. 
This interplay leads to an optimal $\vartheta$ around $40^\circ$. 
In the $\mathrm{SA}$ scenario, since each UAV adjusts its beam toward its intended user exactly, the association probability is not influenced by $\vartheta$ and the coverage probability monotonically decreases with $\vartheta$. 
Moreover, regarding the tendency of coverage probability and CSP variance, we have the same observation as in Fig.~\ref{fig:5b}.

\subsection{Impact of Environment}

\begin{figure}[!t]
    \centering
    \includegraphics[width=0.5\textwidth]{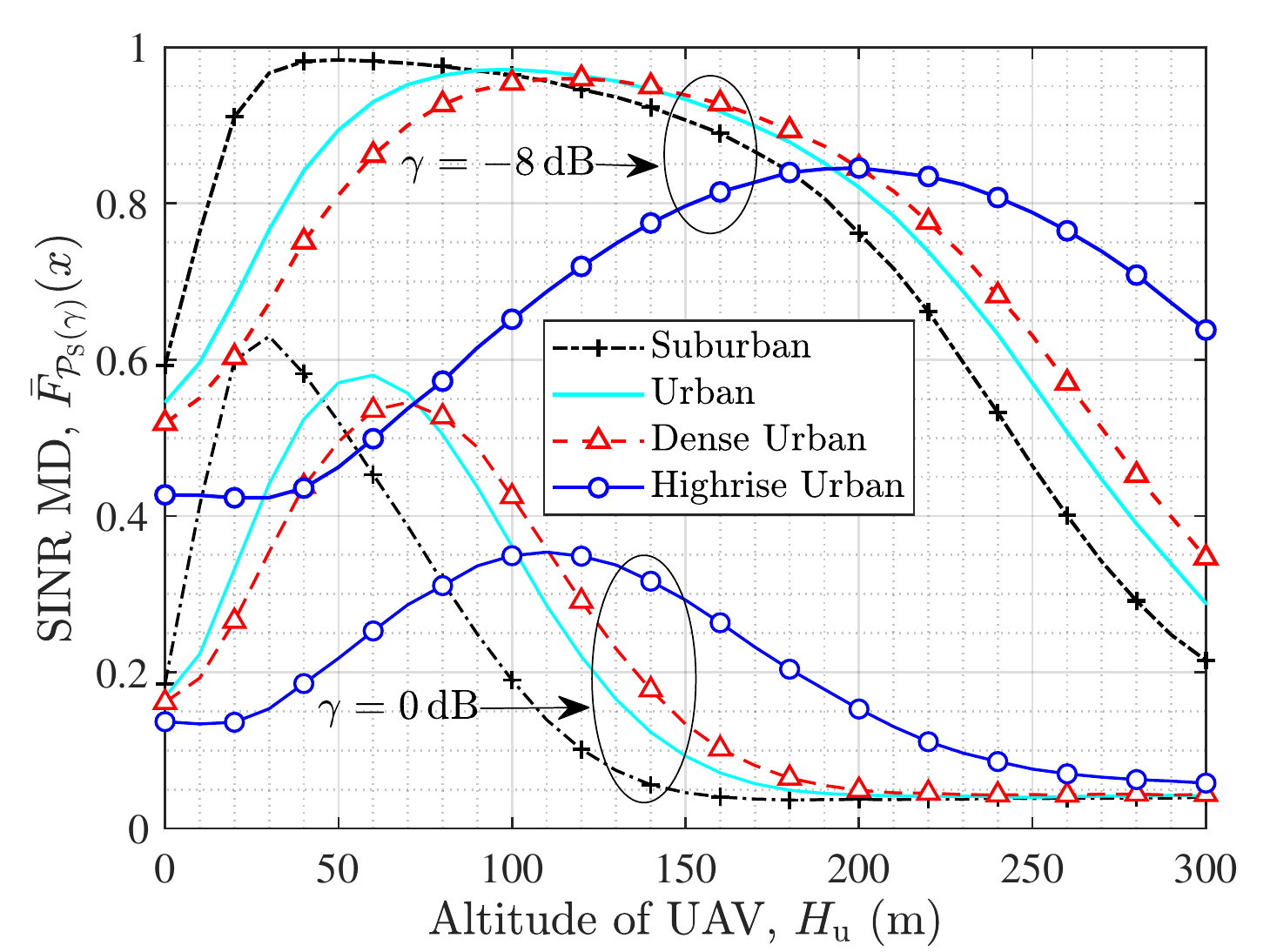}
    \caption{SINR MD as a function of UAV altitude $H$ in different environments when $\varsigma=\mathrm{SA}$ and $x=0.9$. {The parameter $\left\{ \mu _a,\mu _b \right\}$ for suburban, urban, dense urban, and highrise urban environments are given as $\left\{ 4.88,0.43 \right\}$, $\left\{ 9.61,0.16 \right\}$, $\left\{ 11.95,0.14 \right\}$, and $\left\{ 27.23,0.08 \right\}$, respectively \cite{AlHoKandLard14,ZhouDurrGuoYani19}.}}
    \label{fig:8}
\end{figure}

Fig.~\ref{fig:8} shows the SINR MD in different propagation environments. 
One observation is that with the increase of building height and density, i.e., from suburban to highrise urban, the optimal altitude becomes higher. 
This is expected since UAVs have to fly higher to establish LoS links to their serving users in heavily obstructed environments. 
By contrast, in less obstructed environments, UAVs have to fly lower to reduce the detrimental impact of LoS A2G interference.
Another observation is that in suburban environment, SINR MD becomes less sensitive to the change of UAV altitude. 
The reason for this phenomenon is that in heavily obstructed environment, A2G links are more likely to be in NLoS condition, which attenuates the power received from UAVs and thus the coverage performance is dominated by TBSs.

\subsection{Impact of Deployment}

\begin{figure}[!t]
    \centering
    \subfloat[$\lambda_{\mathrm{b}}+\lambda_{\mathrm{u}}=50/\text{km}^2$]{\includegraphics[width=0.5\textwidth]{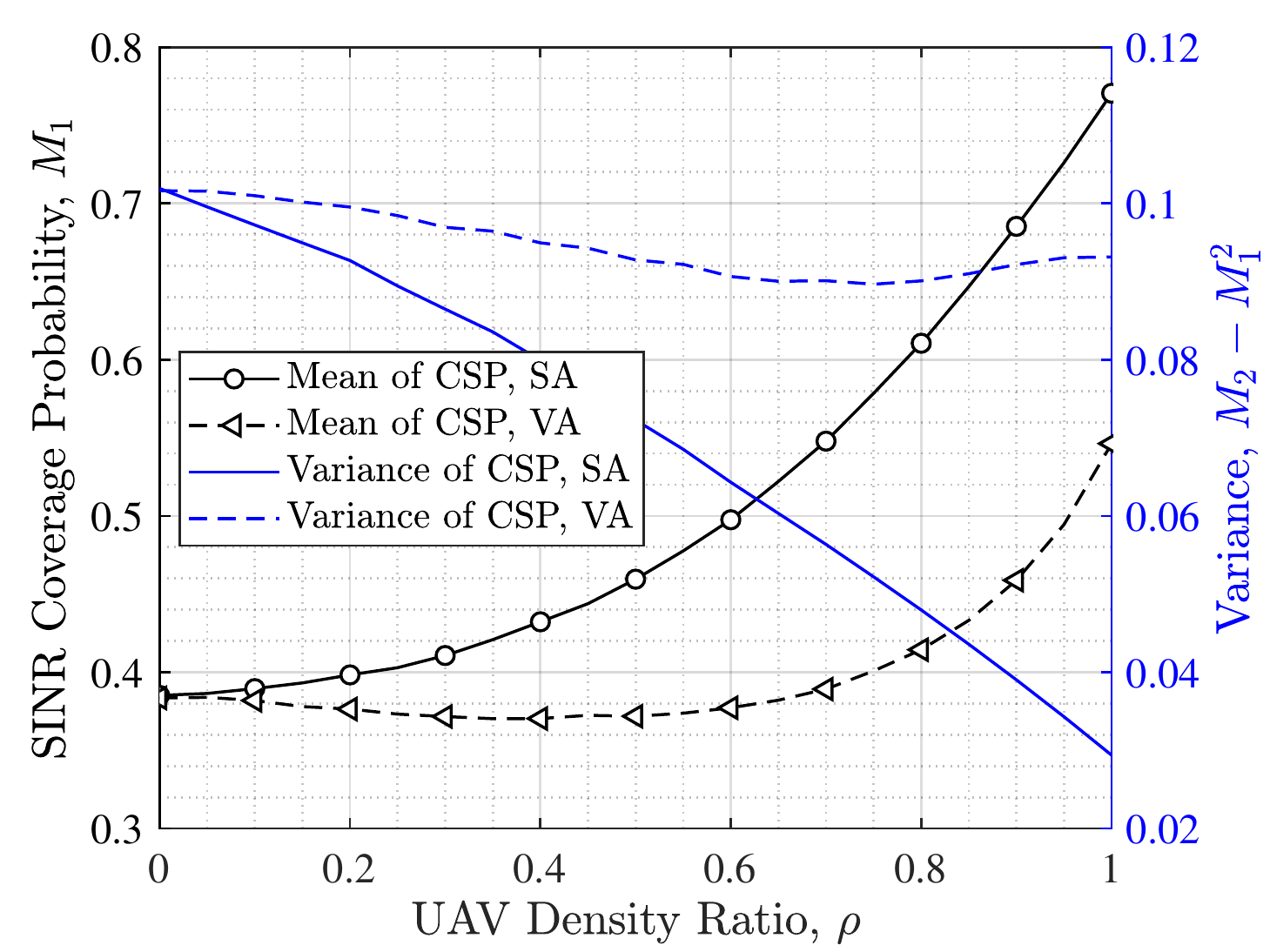}\label{fig:9a}}\hfil
    \subfloat[$\lambda_{\mathrm{b}}+\lambda_{\mathrm{u}}=100/\text{km}^2$]{\includegraphics[width=0.5\textwidth]{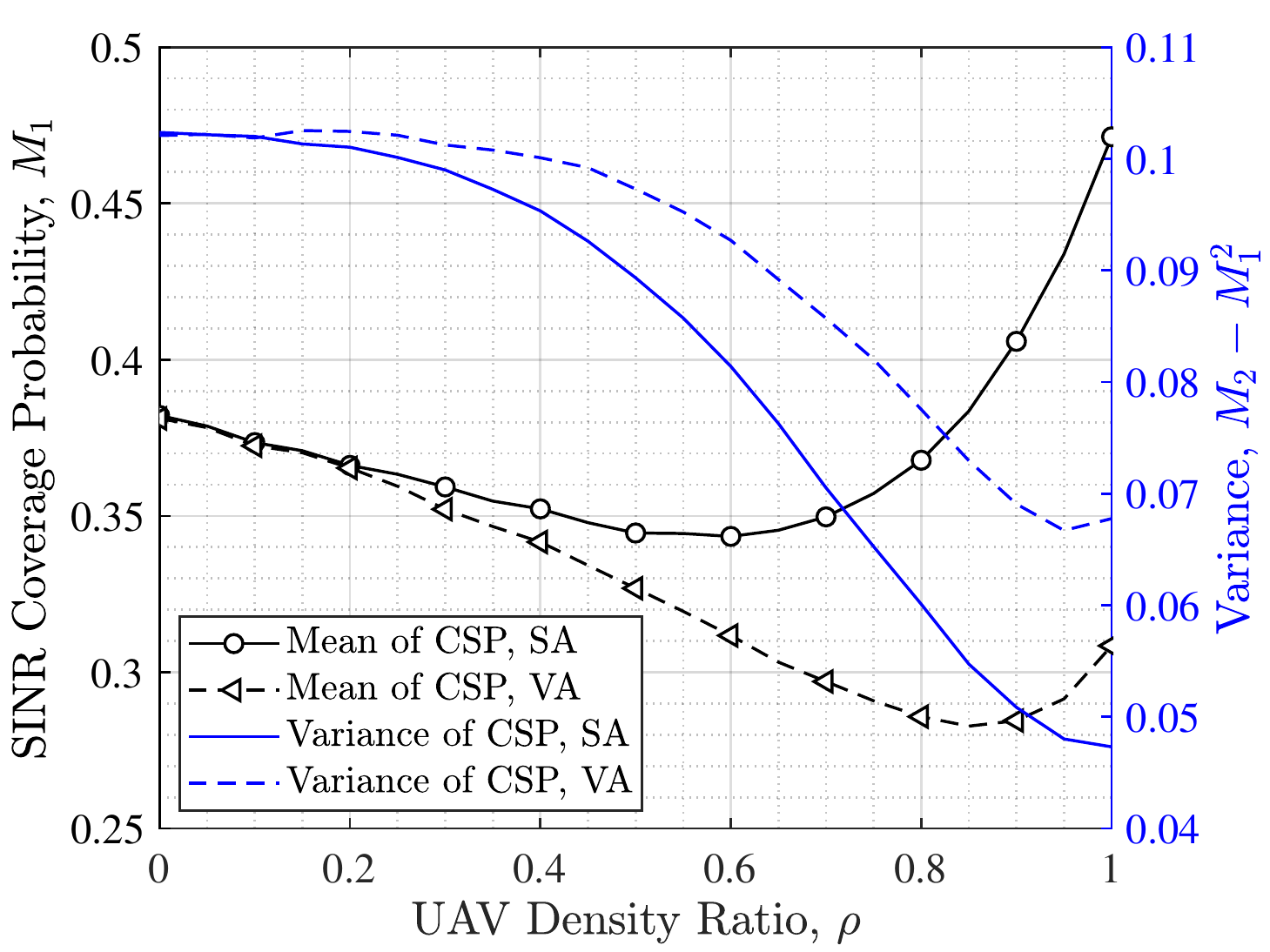}\label{fig:9b}}
    \caption{The mean and variance of CSP as a function of UAV density ratio $\rho$, where $\rho\coloneqq\lambda_{\mathrm{u}}/(\lambda_{\mathrm{b}}+\lambda_{\mathrm{u}})$. Here $\rho=0$ and $\rho=1$ correspond to the TBS-only scenario and UAV-only scenario, respectively.}
    \label{fig:9}
\end{figure}

Fig.~\ref{fig:9} highlights the impacts of UAV density ratio $\rho$ on the mean and variance of CSP, given the total density of TBSs and UAVs. 
It is observed that in sparse networks (e.g., $\lambda_{\mathrm{b}}+\lambda_{\mathrm{u}}=50/\text{km}^2$), replacing TBSs with UAVs helps to increase the coverage performance, which mainly benefits from UAV's capability of establishing LoS transmissions to terrestrial users. 
However, this is not the case for dense networks (e.g., $\lambda_{\mathrm{b}}+\lambda_{\mathrm{u}}=100/\text{km}^2$), where the coverage probability decreases slightly and then starts increasing. 
The reason for this phenomenon is that replacing TBSs with UAVs results in longer G2G communication distance, which may counteract the enhancement of A2G communication. 
Another important observation is that replacing TBSs with UAVs helps to decrease the variance of CSP and potentially improve the fairness among individual links.

\section{Conclusion}\label{sec:conclusion}

This paper provides a fine-grained analysis for UAV-assisted cellular networks based on the SINR MD. 
For the UAV antenna patterns of type $\mathrm{SA}$ and $\mathrm{VA}$, we derive analytical expressions of the probability that a typical user is associated with a specific type of transmitter. 
Considering probabilistic LoS model and general Nakagami fading for A2G links, we derive bounding expressions of CSP and its moments, in which we account for the distance-dependent antenna gain from interfering UAVs. 
Further, the SINR MD is derived by matching its first two moments with standard beta distribution. 
Numerical results validate the tightness of the analytical results and reveal several useful observations: 
$1)$ the widely used uniform OBA assumption underestimates the network performance, which indicates the necessity of incorporating the distance-dependent OBA in analyzing coverage performance and SINR MD; 
$2)$ equipping steerable antennas on UAVs is expected to optimize the network coverage and user fairness simultaneously, while these two metrics need a trade-off for the case of vertical antennas; 
$3)$ using UAVs to replace a certain percent of TBSs is beneficial to the network coverage and user fairness.

\appendices
\numberwithin{equation}{section}

\section{Proof of Theorem~\ref{theo:asso_prob}}\label{app:theo_asso_prob}

{For the typical user, the average power received from its serving TBS/UAV of the $k$-th tier is $l_{k}^{\left( \varsigma \right)}\left( r \right)$. 
Therefore, according to Section~\ref{subsec:2D-strategy}, the association probability of the $k$-th tier is determined by}
\begin{align}
    A_{k}^{\left( \varsigma \right)} &=\mathbb{P} \left( X^{\star}\in \Phi _k \right)\notag
    \\
    &=\mathbb{P} \left( \bigcap_{\ell \in \mathcal{K} \backslash \left\{ k \right\}}{l_{k}^{\left( \varsigma \right)}\left( R_{0,k} \right) \ge {l_{\ell}^{\left( \varsigma \right)}}\left( R_{0,\ell} \right)} \right)\notag
    \\
    &\overset{(a)}{=} \mathbb{P} \left( \bigcap_{\ell \in \mathcal{K} \backslash \left\{ k \right\}}{R_{0,\ell}\ge {\dot{l}_{\ell}^{\left( \varsigma \right)}}\left( l_{k}^{\left( \varsigma \right)}\left( R_{0,k} \right) \right)} \right)\notag
    \\
    &\overset{(b)}{=} \int_0^{\infty}{f_{R_{0,k}}\left( r \right) \prod_{\ell \in \mathcal{K} \backslash \left\{ k \right\}}{\bar{F}_{R_{0,\ell}}\left( {\dot{l}_{\ell}^{\left( \varsigma \right)}}\left( l_{k}^{\left( \varsigma \right)}\left( r \right) \right) \right)}\,\mathrm{d}r},\label{eq:proof_asso_prob_k}
\end{align}
where (a) follows from the monotonic decreasing property of $l_{k}^{\left( \varsigma \right)}(r)$ and (b) from the spatial independence between different tiers. Finally, substituting \eqref{eq:r0k_cdf} and \eqref{eq:r0k_pdf} into \eqref{eq:proof_asso_prob_k} completes the proof.

\section{Proof of Lemma~\ref{lemma:theta_cdf_pdf_cond_l_r}}\label{app:lemma_theta_cdf_pdf_cond_l_r}
{The results for $\mathrm{VA}$ scenario is apparent, and thus we only show the derivation for $\mathrm{SA}$ scenario. 
In Fig.~\ref{fig:2a}, applying the law of cosines to $\angle OP^{\prime}Q$, the length of $OQ$ is formulated by
\begin{align}
    \overline{OQ} &=\sqrt{L^2+t^2-2Lt\cos \left( \pi -\alpha \right)}=\sqrt{L^2+t^2+2Lt\cos \alpha}.
\end{align}
Then, applying the law of cosines to $\angle OPQ$ yields
\begin{align}
    \cos \Theta =\cos\angle OPQ =\frac{\overline{OP}^2+\overline{PQ}^2-\overline{OQ}^2}{2\overline{OP}\cdot\overline{PQ}} =\frac{H_{\mathrm{u}}^2-Lt\cos \alpha}{\sqrt{\left( H_{\mathrm{u}}^2+L^2 \right) \left( H_{\mathrm{u}}^2+t^2 \right)}}.\label{eq:cosTheta}
\end{align}
Note that $\Theta$ is minimized to $\theta _{\min}$ when $\alpha=\pi$ and is maximized to $\theta _{\max}$ when $\alpha=0$. 
When $\theta \in \left[ \theta _{\min},\theta _{\max} \right]$, based on \eqref{eq:cosTheta}, we have
\begin{align}
    F_{\left. \Theta \right|L,t}\left( \theta \right) &=\mathbb{P} \left( \left. \Theta \le \theta \right|L,t \right) \notag
    \\
    &=\mathbb{P} \left( \left. \cos \Theta \ge \cos \theta \right|L,t \right) \notag
    \\
    &=\mathbb{P} \left( \left. \frac{H_{\mathrm{u}}^2-Lt\cos \alpha}{\sqrt{\left( H_{\mathrm{u}}^2+L^2 \right) \left( H_{\mathrm{u}}^2+t^2 \right)}}\ge \cos \theta \right|L,t \right) \notag
    \\
    &=\mathbb{P} \left( \left. \cos \alpha \le \frac{H_{\mathrm{u}}^2-\sqrt{\left( H_{\mathrm{u}}^2+L^2 \right) \left( H_{\mathrm{u}}^2+t^2 \right)}\cos \theta}{Lt} \right|L,t \right).
\end{align}}
Since $\Phi_k$ is isotropic in the $xy$-plane, we conclude that $\alpha$ is uniformly distributed in $[0,2\pi]$ and \eqref{eq:theta_cdf_cond_l_r} is then obtained. Finally, taking the derivative of $F_{\left. \Theta \right|L,t}\left( \theta \right)$ with respect to $\theta$ yields at \eqref{eq:theta_pdf_cond_l_r}.

\section{Proof of Theorem~\ref{theo:mbk_sa_va}}\label{app:theo_mbk_sa_va}
{Since the bound in \eqref{eq:csp_k_expression} is tight, the moment $M_{b|k}^{\left( \varsigma \right)}\left( \gamma \right)$, $b\in\mathbb{N}$ is approximated by}
\begin{align}
    M_{b|k}^{\left( \varsigma \right)}\left( \gamma \right) &=\mathbb{E} _{\Phi _{\mathrm{b}},\Phi _{\mathrm{u}}}\left[ \mathcal{P} _{\mathrm{s}|k}\left( \gamma |\Phi _{\mathrm{b}},\Phi _{\mathrm{u}} \right) ^b \right] \notag
    \\
    &\approx \mathbb{E} _{\Phi _{\mathrm{b}},\Phi _{\mathrm{u}}}\left[ \left( 1-\mathbb{E} _{\mathsf{H},G}\left[ \left( 1-\exp \left( -\frac{\gamma \phi _k\left( N_0+\sum_{\ell \in \mathcal{K}}{I_{\ell}} \right)}{l_{k}^{\left( \varsigma \right)}\left( Y_{0,k} \right)} \right) \right) ^{\mathsf{M}_k} \right] \right) ^b \right] \notag
    \\
    &{\coloneqq\tilde{M}_{b|k}^{\left( \varsigma \right)}\left( \gamma \right).} \label{eq:proof_mb_1}
\end{align}
{Since $f\left( x \right) =\left( 1-x \right) ^b$ is convex in $[0,1]$, by invoking Jensen's inequality, $\tilde{M}_{b|k}^{\left( \varsigma \right)}\left( \gamma \right)$ is upper bounded by} 
\begin{align}
    {\tilde{M}_{b|k}^{\left( \varsigma \right)}\left( \gamma \right)} &\leq \mathbb{E} _{\Phi _{\mathrm{b}},\Phi _{\mathrm{u}}}\mathbb{E} _{\mathsf{H},G}\left[ \left( 1-\left( 1-\exp \left( -\frac{\gamma \phi _k\left( N_0+\sum_{\ell \in \mathcal{K}}{I_{\ell}} \right)}{l_{k}^{\left( \varsigma \right)}\left( Y_{0,k} \right)} \right) \right) ^{\mathsf{M}_k} \right) ^b \right] \notag
    \\
    &\overset{(a)}{=}\sum_{n=0}^b{\sum_{m=0}^{\mathsf{M} _kn}{\binom{b}{n}\binom{\mathsf{M} _kn}{m}\left( -1 \right) ^{n+m}\mathcal{T} _{k,m}\left( \gamma \right)}}, \label{eq:proof_mb_2}
\end{align}
where (a) follows from the binomial theorem and $\mathcal{T} _{k,m}\left( \gamma \right)$ is 
\begin{align}
    \mathcal{T} _{k,m}\left( \gamma \right) &=\mathbb{E} _{\Phi _{\mathrm{b}},\Phi _{\mathrm{u}}}\mathbb{E} _{\mathsf{H},G}\left[ \exp \left( -\frac{m\gamma \phi _k}{l_{k}^{\left( \varsigma \right)}\left( Y_{0,k} \right)}\left( N_0+\sum_{\ell \in \mathcal{K}}{I_{\ell}} \right) \right) \right] \notag
    \\
    &=\mathbb{E} _{\Phi _{\mathrm{b}},\Phi _{\mathrm{u}}}\mathbb{E} _{\mathsf{H},G}\left[ \exp \left( -\frac{m\gamma \phi _kN_0}{l_{k}^{\left( \varsigma \right)}\left( Y_{0,k} \right)} \right) \prod_{\ell \in \mathcal{K}}{\exp \left( -\frac{m\gamma \phi _kI_{\ell}}{\zeta _kY_{0,k}^{-\alpha _k}} \right)} \right] \notag
    \\
    &{\overset{(b)}{=}\int_0^{\infty}{f_{Y_{0,k}^{\left( \varsigma \right)}}\left( y \right) \exp \left( -\frac{m\gamma \phi _kN_0}{l_{k}^{\left( \varsigma \right)}\left( y \right)} \right) \prod_{\ell \in \mathcal{K}}{\mathcal{L} _{I_{\ell}}\left( \frac{m\gamma \phi _k}{l_{k}^{\left( \varsigma \right)}\left( y \right)} \right)}\mathrm{d}y}}, \label{eq:proof_mb_3}
\end{align}
{where (b) is from the independence among the point processes of different tiers and $\mathcal{L} _{I_{\ell}}\left( \cdot \right)$ is the Laplace transform of $I_{\ell}$. 
Specifically, recalling \eqref{eq:Ik_define}, $\mathcal{L} _{I_{\ell}}\left( \cdot \right)$ is formulated by 
\begin{align}
    \mathcal{L} _{I_{\ell}}\left( s \right) &\coloneqq \mathbb{E} \left[ \exp \left( -sI_{\ell} \right) \right] \notag
    \\
    &=\mathbb{E} _{\Phi _{\ell},\mathsf{H},G}\left[ \exp \left( -s\sum_{X\in \Phi _{\ell}\backslash \left\{ X^{\star} \right\}}{P_{\ell}G_X\mathsf{H}_X\kappa _{\ell}\left\| X \right\| ^{-\alpha _{\ell}}} \right) \right] \notag
    \\
    &=\mathbb{E} _{\Phi _{\ell}}\left[ \prod_{X\in \Phi _{\ell}\backslash \left\{ X^{\star} \right\}}{\mathbb{E} _{\mathsf{H},G}\left[ \exp \left( -sP_{\ell}G_X\mathsf{H}_X\kappa _{\ell}\left\| X \right\| ^{-\alpha _{\ell}} \right) \right]} \right] \notag
    \\
    &\overset{(a)}{=}\mathbb{E} _{\Phi _{\ell}}\left[ \prod_{X\in \Phi _{\ell}\backslash \left\{ X^{\star} \right\}}{\mathcal{M} _{\mathsf{H}_{\ell}}\left( sh_{\ell}^{\left( \varsigma \right)}\left( \left\| X \right\| \right) \right)} \right], \label{eq:proof_mb_4}
\end{align}
where (a) follows by taking the expectation over $\Theta$ using \textbf{Lemma~\ref{lemma:theta_cdf_pdf_cond_l}} and using the moment generating function of the gamma random variable $\mathsf{H}_{\ell}$, i.e., $\mathcal{M} _{\mathsf{H}_{\ell}}\left( s \right) =\left( 1+s/\mathsf{M}_{\ell} \right) ^{-\mathsf{M}_{\ell}}$. 
Then, leveraging the probability generating functional (PGFL) of PPP $\Phi_{\ell}$ \cite{BlasHaenKeelMukh18BK} yields 
\begin{align}
    \mathcal{L} _{I_{\ell}}\left( s \right) &=\exp \left\{ -\int_{\chi _{k,\ell}^{\left( \varsigma \right)}\left( y \right)}^{\infty}{\left[ 1-\mathcal{M} _{\mathsf{H}_{\ell}}\left( sh_{\ell}^{\left( \varsigma \right)}\left( r \right) \right) \right] \bar{\lambda}_{\ell}\left( r \right) \,\mathrm{d}r} \right\}. \label{eq:proof_mb_5}
\end{align}
Finally, substituting \eqref{eq:y0k_pdf} and \eqref{eq:proof_mb_2} into \eqref{eq:mbk_to_mb} completes the proof.}

\bibliographystyle{IEEEtran}

\end{document}